%% file: main.tex
\documentclass[twocolumn, tighten]{aastex62}
\usepackage{amsmath}

\input{results}

\defcitealias{dehaan16}{dH16}
\defcitealias{dietrich19}{D19}
\defcitealias{schrabback18b}{S18}

\renewcommand{\vec}[1]{\mbox{\boldmath$#1$}}

\newcommand{\hst}{\textit{Hubble Space Telescope}}
\newcommand{\planck}{\textit{Planck}}
\newcommand{\chandra}{\textit{Chandra}}
\newcommand{\spitzer}{\textit{Spitzer}}
\newcommand{\mgas}{\ensuremath{M_\mathrm{gas}}}
\newcommand{\fgas}{\ensuremath{f_\mathrm{gas}}}
\newcommand{\yx}{\ensuremath{Y_\mathrm{X}}}
\newcommand{\mwl}{\ensuremath{{M_\mathrm{WL}}}}

\newcommand{\asz}{\ensuremath{{A_\mathrm{SZ}}}}
\newcommand{\bsz}{\ensuremath{{B_\mathrm{SZ}}}}
\newcommand{\csz}{\ensuremath{{C_\mathrm{SZ}}}}
\newcommand{\sigmalnzeta}{\ensuremath{{\sigma_{\ln \zeta}}}}
\newcommand{\ayx}{\ensuremath{{A_{\yx}}}}
\newcommand{\byx}{\ensuremath{{B_{\yx}}}}
\newcommand{\cyx}{\ensuremath{{C_{\yx}}}}
\newcommand{\sigmalnyx}{\ensuremath{{\sigma_{\ln \yx}}}}
\newcommand{\amg}{\ensuremath{{A_{M_\mathrm{g}}}}}
\newcommand{\bmg}{\ensuremath{{B_{M_\mathrm{g}}}}}
\newcommand{\cmg}{\ensuremath{{C_{M_\mathrm{g}}}}}
\newcommand{\sigmalnmgas}{\ensuremath{{\sigma_{\ln M_\mathrm{g}}}}}
\newcommand{\bwl}{\ensuremath{{b_\mathrm{WL}}}}
\newcommand{\sigmawl}{\ensuremath{{\sigma_\mathrm{WL}}}}

\newcommand{\Om}{\ensuremath{\Omega_\mathrm{m}}}
\newcommand{\Omnuhh}{\ensuremath{\Omega_\nu h^2}}
\newcommand{\Ombhh}{\ensuremath{\Omega_\mathrm{b} h^2}}
\newcommand{\sig}{\ensuremath{\sigma_8}}
\newcommand{\sigz}{\ensuremath{\sigma_8(z)}}
\newcommand{\LCDM}{\ensuremath{\Lambda\mathrm{CDM}}}
\newcommand{\nuLCDM}{\ensuremath{\nu\Lambda\mathrm{CDM}}}
\newcommand{\wCDM}{\ensuremath{w\mathrm{CDM}}}
\newcommand{\nuwCDM}{\ensuremath{\nu w\mathrm{CDM}}}
\newcommand{\sigOmtwo}{\ensuremath{\sigma_8(\Omega_\mathrm m/0.3)^{0.2}}}
\newcommand{\sigOmfive}{\ensuremath{\sigma_8(\Omega_\mathrm m/0.3)^{0.5}}}
\newcommand{\sumMnu}{\ensuremath{\sum m_\nu}}

\shorttitle{SPT-SZ Cluster Cosmology with Weak-Lensing Mass Calibration}
\shortauthors{Bocquet et al.}

\accepted{April 2019}
\submitjournal{\apj}

\begin{document}

\title{Cluster Cosmology Constraints from the 2500~deg$^2$ SPT-SZ Survey:\\
Inclusion of Weak Gravitational Lensing Data from Magellan and the \textit{Hubble Space Telescope}}

\input{authors}
\email{sebastian.bocquet@physik.lmu.de}

\begin{abstract}
\input{abstract}
\end{abstract}


\section{Introduction}

Measurements of the abundance of galaxy clusters have become an important part
of the cosmological toolkit. Galaxy clusters and their associated dark matter
halos trace the highest and therefore rarest peaks in the matter density field
on megaparsec scales. To obtain cosmological constraints, one confronts the
predicted halo abundance, the \emph{halo mass function} (HMF), which is provided
by numerical cosmological simulations, with the observations. The key challenge
is to accurately describe the relation between halo mass in the simulations and
the observable quantities. The cluster abundance essentially constrains the
parameter combination $\sig(\Om/0.3)^\alpha$, where \sig\ is the root mean
square fluctuation in the linear matter density field on $8$~Mpc$/h$ scales at
$z=0$ and $\alpha$ is of the order of about $0.2-0.4$ depending on survey
specifics. Measuring the cluster abundance over a range of redshifts enables
constraints on the cosmic expansion and structure formation histories. This
probe can therefore be used to challenge the paradigms of a cosmological
constant and of General Relativity, and, when analyzed jointly with measurements
of primary anisotropies in the cosmic microwave background (CMB), to measure the
sum of neutrino masses \citep[for reviews, see, e.g.,][]{allen11, kravtsov12}.

Cosmological analyses have been performed using samples of galaxy clusters
constructed from their observed galaxy populations \citep[e.g.,][]{rykoff16},
their X-ray emission \citep[e.g.,][]{vikhlinin09, mantz10a}, and their
millimeter-wave signal \citep[e.g.,][]{bleem15b, planck15-24, hilton18}. The
latter is dominated by the thermal Sunyaev-Zel'dovich effect
\citep[SZ;][]{sunyaev72} which arises when CMB photons scatter off hot electrons
in the intracluster medium (ICM). The surface brightness of the SZ effect is
independent of cluster redshift, and high-resolution mm-wave surveys can
therefore be used to construct clean and essentially mass-limited catalogs out
to the highest redshifts at which clusters exist. This makes SZ-selected cluster
samples particularly suited for studying the evolution of scaling relations and
the growth of cosmic structure over a significant fraction of the age of the
Universe.

In this paper, we present an analysis of the 2500~deg$^2$ SPT-SZ survey cluster
sample that is enabled by optical weak gravitational lensing (WL) data for
SPT-SZ clusters. The WL dataset consists of two subsamples: i) 19 clusters at
intermediate redshifts $0.28<z<0.63$, with ground-based Magellan/Megacam
imaging, referred to as the ``Megacam sample'' hereafter \citep[][hereafter
D19]{dietrich19}; ii) 13 clusters at higher redshifts $0.58<z<1.13$ observed
with the \hst, referred to as the ``HST sample'' hereafter \citep[][hereafter
S18]{schrabback18b}. Using these WL data in our analysis has two main
advantages: i) it removes the need to rely on external calibrations of the
observable--mass relations, ii) our analysis now only considers clusters that
are actually part of the SPT-SZ sample which ensures a fully self-consistent
handling of selection effects.

This work represents an improvement over the first cosmological analysis of the
SZ-selected cluster sample from the full 2500~deg$^2$ SPT-SZ survey
\citep[][hereafter dH16]{dehaan16}, where we combined the cluster number counts
in SZ significance and redshift with X-ray \yx\ follow-up \citep[\yx\ is the
product of X-ray gas mass \mgas\ and temperature $T_\mathrm{X}$;][]{kravtsov06a}
of 82 clusters. The \citetalias{dehaan16} analysis relied on external, WL-based
calibrations of the normalization of the \yx--mass relation and the assumption
that its evolution in mass and redshift follows the self-similar expectation
within some uncertainty (5\% and 50\% uncertainty at $1\sigma$ on the parameters
of the mass and redshift evolution, respectively).

As already mentioned, the key challenge in cluster cosmology is to robustly
model the relation between the observables (SZ signal, WL shear profiles, X-ray
\yx\ measurements) and the underlying, unobserved halo mass, which is the link
to the predicted HMF.\footnote{Although some of the observables carry
cosmological dependences themselves, we seek to constrain cosmology primarily
through its impact on the HMF.} Our modeling assumptions are:
\begin{itemize}
\item The relation between true halo mass and the observed WL signal, and the
scatter around this mean relation are well understood, with systematic
uncertainties at the few percent level. We use numerical simulations to account
for the effects of halo triaxiality, miscentering, and correlated large-scale
structure along the line of sight. Uncorrelated large-scale structure along the
line of sight is accounted for in a semi-analytic approach (Megacam sample) and
via simulated cosmic shear fields (HST sample). For the Megacam sample, the
systematic limit in mass is $5.6\%$ \citepalias{dietrich19}, and it is
$9.2-9.4\%$ for the HST sample \citepalias{schrabback18b}.
\item The mean relations between true halo mass and the SZ and X-ray observables
are described by power-law relations in mass and the dimensionless Hubble
parameter $E(z)\equiv H(z)/H_0$. This functional form is motivated by the
self-similar model \citep[evolution assuming only gravity is at
play;][]{kaiser86b} and confirmed using numerical $N$-body and hydrodynamical
simulations \citep[e.g.,][]{vanderlinde10,dehaan16,gupta17a}. However, we do not
assume any a-priori knowledge of the parameters in these relations and allow for
departures  from self-similarity by marginalizing over wide priors.
\item The intrinsic scatter in the SZ and X-ray observable--mass relations is
described by lognormal distributions (with a-priori unknown width). The
scatter among all three observables may be correlated, and we marginalize over
the correlation coefficients.
\end{itemize}

This paper is organized as follows: In Section 2, we provide an overview of the
cluster dataset and of external cosmological data used in the analysis. We
describe our analysis method in Section 3. In Section 4, we present our
constraints on scaling relations and cosmology. We summarize our findings in
Section 5 and provide some outlook. Further robustness tests are discussed in
the Appendices~\ref{sec:appendixneutrino}--\ref{sec:appendixwcdm}.

Throughout this work we assume spatially flat cosmological models. Cluster
masses are referred to as $M_{\Delta\mathrm c}$, the mass enclosed within a
sphere of radius $r_\Delta$, in which the mean matter density is equal to
$\Delta$ times the critical density. The critical density at the cluster's
redshift is $\rho_\mathrm{crit}(z) = 3H^2(z)/8\pi G$, where $H(z)$ is the Hubble
parameter. We refer to the vector of cosmology and scaling relation parameters
as $\vec p$.

All quoted constraints correspond to the mean and the shortest $68\%$ credible
interval, computed from the MCMC chains using a Gaussian kernel density
estimator.\footnote{\url{https://github.com/cmbant/getdist}} All
multi-dimensional posterior probability plots show the 68\% and 95\% contours.
We use standard notation for statistical distributions, i.e. the normal
distribution with mean $\vec{\mu}$ and covariance matrix $\vec\Sigma$ is
written as $\mathcal{N}(\vec{\mu}, \vec\Sigma)$, and $\mathcal{U}(a, b)$
denotes the uniform distribution on the interval $[a, b]$.

\section{Data}

\begin{figure}
  \includegraphics[width=\columnwidth]{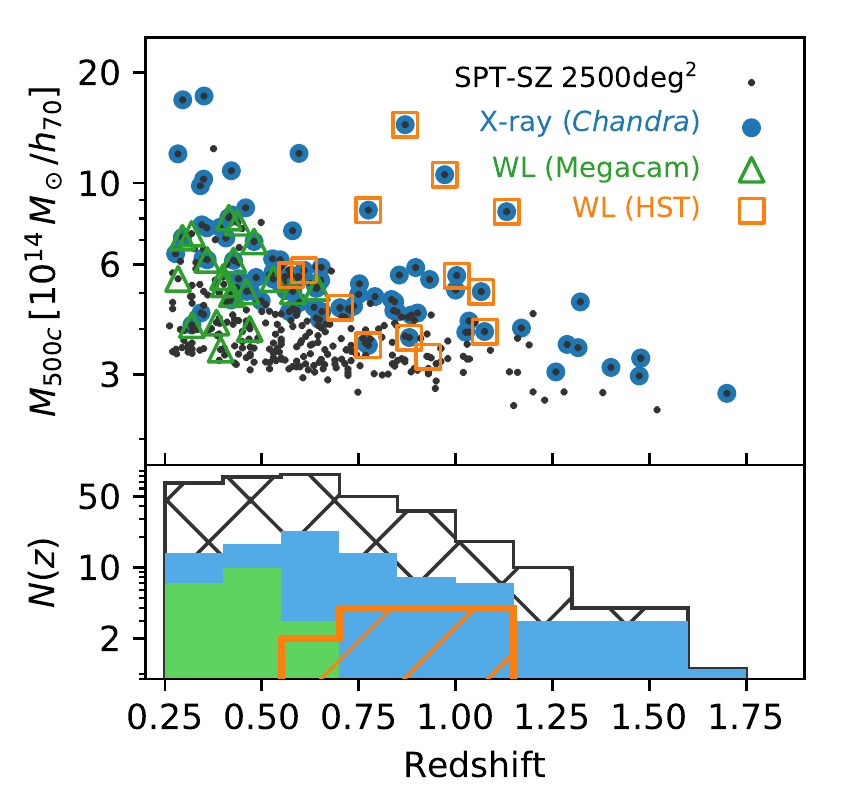}
  \caption{The SPT-SZ 2500~deg$^2$ cluster cosmology sample, selected to have
  redshift $z>0.25$ and detection significance $\xi>5$.
  \emph{Top panel:} The distribution of clusters in redshift and mass (assuming
  a fiducial observable--mass relation). Black points show the full sample, blue
  dots mark those 89 clusters for which X-ray follow-up data from \chandra\ are
  available, and green triangles (orange squares) mark those 19 with
  Magellan/Megacam (13 with the \hst) WL follow-up data.
  \emph{Bottom panel:} Histograms with the same color coding. While the X-ray
  follow-up dataset covers the entire redshift range, the WL follow-up covers
  $0.25<z\lesssim1.1$.}
  \label{fig:sample}
\end{figure}

The cluster cosmology sample from the 2500~deg$^2$ SPT-SZ survey consists of 365
candidates of which 343 are optically confirmed and have redshift measurements.
X-ray follow-up measurements with \chandra\ are available for 89 clusters, and
WL shear profiles are available for 19 clusters from ground-based observations
with Magellan/Megacam and for 13 clusters observed from space with the \hst\
(see Fig.~\ref{fig:sample}).

\subsection{The SPT-SZ 2500 deg$^2$ Cluster Sample}
\label{sec:SPTsample}

The South Pole Telescope (SPT) is a 10~m telescope located within 1~km of the
geographical South Pole \citep{carlstrom11}. The $\sim$1~arcmin resolution and
1 degree field of view are well suited for a survey of rare, high-mass clusters
from a redshift of $z\geq0.2$ out to the highest redshifts where they exist.
From 2007 to 2011, the telescope was configured to observe with the SPT-SZ
camera in three millimeter-wave bands (centered at 95, 150, and 220~GHz). The
majority of this period was spent on the SPT-SZ survey, a contiguous
2500~deg$^2$ area within the boundaries 20h~$\leq$~R.A.~$\leq$~7h and
$-65^\circ\leq\mathrm{Dec.}\leq-40^\circ$. The survey achieved a fiducial depth
of $\leq18~\mu$K-arcmin in the 150~GHz band.

Galaxy clusters are detected via their thermal SZ signature in the 95 and
150~GHz maps. These maps are created using time-ordered data processing and
map-making procedures equivalent to those described in \cite{vanderlinde10,
reichardt13}. Galaxy clusters are extracted using a multi-scale matched-filter
approach \citep{melin06} applied to the multi-band data as described in
\cite{williamson11,reichardt13}.

We use the same SPT-SZ cluster sample that was analyzed in
\citetalias{dehaan16}. Namely, this \textit{cosmological sample} is a subset of
the full SPT-SZ cluster sample presented in \cite{bleem15b}, restricted to
redshifts $z>0.25$ and detection significances $\xi>5$. This cosmological sample
has an expected and measured purity of 95\% \citep{bleem15b}. For clusters at
redshifts below $z=0.25$, confusion with primary CMB fluctuations changes the
scaling of the $\xi$--mass relation.

\begin{figure}
  \includegraphics[width=\columnwidth]{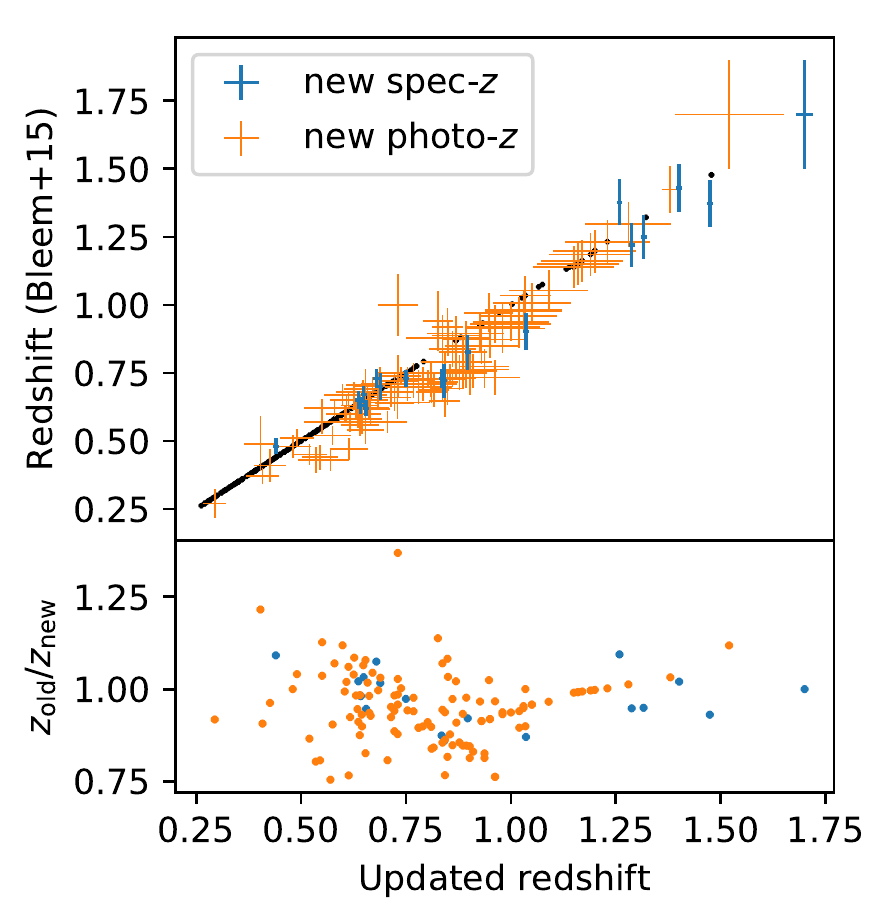}
  \caption{Updates in cluster redshifts since the publication of the SPT-SZ
  cluster catalog \citep{bleem15b}.
  \emph{Top panel}: Original redshifts plotted against the updated ones. Black
  points show unchanged redshifts (without error bars for ease of presentation),
  orange error bars show updated photometric redshifts, and blue error bars show
  new spectroscopic measurements.
  \emph{Bottom panel:} Changes in redshifts; we omit unchanged redshifts and all
  error bars. Orange points show the change in photo-$z$s, blue points show
  changes due to new spec-$z$ measurements.}
  \label{fig:redshifts}
\end{figure}

We have improved the cluster redshift estimates from the original values
provided in \citet{bleem15b} to incorporate both new spectroscopic measurements
\citep[][Mantz et al., in prep.]{bayliss16,khullar19}, two updated high-redshift
photo-$z$ measurements with \hst\ \citep{strazzullo19}, and improved photometric
measurements. These improved photometric redshifts are enabled both via the
recalibration of our \spitzer\ redshift models using the new spectroscopic data
and by the use of optical data from the Parallel Imager for Southern Cosmology
Observations (PISCO), a new imager installed on the Magellan/Clay telescope at
Las Campanas Observatory \citep{stalder14}. PISCO---with a fast ($\sim20$~s)
readout, 9~arcmin field-of-view, and simultaneous 4-band (\textit{griz}) imaging
capability---is optimized for efficient characterization of clusters and other
systems identified from external surveys. As part of further efforts to
characterize the SPT-SZ cluster sample, we have obtained approximately uniform
PISCO imaging for the majority of the previously confirmed SPT-SZ clusters.
Notably, this deeper optical data has allowed less constraining infrared-driven
redshift estimates from \spitzer\ to be replaced by more robust estimates based
on optical red-sequence techniques for a significant number of clusters in the
range $0.8\lesssim z\lesssim1$. As a consequence, while the improved data and
model calibration results in small changes in redshift estimates for systems at
$z\lesssim0.8$ and $z\gtrsim1$, at intermediate redshifts, replacing
infrared-driven redshifts with more robust optical estimates leads to up to
$1.5\sigma$ systematic shifts, see Fig.~\ref{fig:redshifts}. We will briefly
come back to this issue in Section~\ref{sec:wCDM}.

\subsection{X-ray Measurements}
\label{sec:Xraydata}

We use X-ray measurements for a subsample of 89 clusters. Eighty-one of these
were also used in our previous cosmological analysis \citepalias{dehaan16}. We
decided not to use the X-ray data available for SPT-CL~J0142-5032 because of its
large measurement error in temperature exceeding 40\%. This has a negligible
impact on our results. Most of those X-ray measurements were originally
presented in \cite{mcdonald13}, and they were largely acquired through a
\chandra\ X-ray Visionary Project (PI: Benson). This sample is now supplemented
with observations of 8 high-redshift $z>1.2$ clusters \citep{mcdonald17}. We
refer the reader to these references for the details of the X-ray analysis.

The X-ray data products entering this analysis are: i) lookup tables of the
total gas mass, \mgas\ within an outer radius ranging from $80-2000$~kpc
(calculated using a fiducial cosmology), allowing interpolation of \mgas\ within
any realistic value of $r_{500}$, and ii) spectroscopic temperatures,
$T_\mathrm{X}$, in the $0.15-1.0 \, r_{500}$ aperture. All X-ray measurements
were re-made for this work using the \chandra\ calibration CALDB~v4.7.7. Note
that this calibration does not change the results from \citetalias{dehaan16}.

\subsection{Weak Gravitational Lensing Data}

We use WL measurement for 32 clusters in our
sample. Of these, 19 were observed with Magellan/Megacam at redshifts
$0.29 \leq z \leq 0.69$ \citepalias{dietrich19}, and 13 at redshifts
$0.576 \leq z \leq 1.132$ with the Advanced Camera for Surveys onboard the
\hst\ \citepalias{schrabback18b}. Details on the data reduction and
analysis methods can be found in these works.

The data products from these works used in our analysis are the reduced
tangential shear profiles in angular coordinates, corrected for contamination by
cluster galaxies, and the estimated redshift distributions of the selected
source galaxies. These are the observable quantities, which are independent from
cosmology, whereas mass estimates or shear profiles in physical coordinates
depend on cosmology through the redshift distance relation and the cosmology
dependence of the NFW profile. Our approach ensures a clean separation between
the actual measurements and their modeling.

\subsection{External Cosmological Data Sets}

In addition to our cluster dataset, we will also consider external cosmological
probes. We use measurements of primary CMB anisotropies from \textit{Planck} and
focus on the TT+lowTEB data combination from the 2015 analysis
\citep{planck15-13}. We use angular diameter distances as probed by Baryon
Acoustic Oscilations (BAO) by the 6dF Galaxy Survey \citep{beutler11}, the SDSS
Data Release~7 Main Galaxy Sample \citep{ross15}, and the BOSS Data Release~12
\citep{alam17}. We also use measurements of luminosity distances from Type Ia
supernovae from the Pantheon sample \citep{scolnic18}.

\section{Analysis Method}

In this section, we present the observable--mass relations, the likelihood
function, and the priors adopted. Fig.~\ref{fig:pipeline} shows a flowchart of
the analysis pipeline. The data and likelihood code will be made publicly
available.

\subsection{Observable--mass Relations}
\label{sec:massobsrel}

We consider three cluster mass proxies: the unbiased SZ significance $\zeta$,
the X-ray \yx, and the WL mass $M_\mathrm{WL}$. We parametrize the mean
observable--mass relations as
\begin{equation}
  \begin{split}
    \langle\ln\zeta\rangle =& \ln\asz + \bsz \ln\left(\frac{M_{500c}\, h_{70}}{4.3\times 10^{14}M_\odot}\right) \\
    &+ \csz \ln\left(\frac{E(z)}{E(0.6)}\right) \label{eq:zetaM}
  \end{split}
\end{equation}
\begin{equation}
  \begin{split}
    \ln\left(\frac{M_{500c}\, h_{70}}{8.37\times 10^{13}M_\odot}\right) =& \ln\ayx + \byx\langle\ln\yx\rangle \\
    &+ \byx\ln\left(\frac{h_{70}^{5/2}}{3\times10^{14}M_\odot \mathrm{keV}}\right) \\
    &+ \cyx\ln E(z) \label{eq:YxM}
  \end{split}
\end{equation}
\begin{equation}
\langle\ln M_\mathrm{WL}\rangle = \ln b_\mathrm{WL} +\ln M_{500c}. \label{eq:MwlM}
\end{equation}
The $\zeta$--mass and \yx--mass relations are equivalent to the ones adopted in
\citetalias{dehaan16}, except for replacing $h/0.72$ by $h_{70}$ in \yx--mass.

The intrinsic scatter in $\ln\zeta$, $\ln\yx$, and $\ln M_\mathrm{WL}$ at fixed
mass and redshift is described by normal distributions with widths
\sigmalnzeta, \sigmalnyx, and \sigmawl. These widths are assumed to be
independent of mass and redshift. Note that the parameters \sigmalnzeta\ and
\sigmalnyx\ have been called $D_\mathrm{SZ}$ and $D_\mathrm{X}$ in some previous
SPT publications. We allow for correlated scatter between the SZ, X-ray, and WL
mass proxies as described by the covariance matrix
\begin{equation}
  \begin{split}
    &\vec\Sigma_\text{multi-obs} = \\
    &\begin{pmatrix}
      \sigma_{\ln\zeta}^2 & \rho_\mathrm{SZ-WL}\sigmalnzeta\sigmawl & \rho_\mathrm{SZ-X}\sigmalnzeta\sigmalnyx \\
      \rho_\mathrm{SZ-WL}\sigmalnzeta\sigmawl & \sigma_\mathrm{WL}^2 & \rho_\mathrm{WL-X}\sigmawl\sigmalnyx \\
      \rho_\mathrm{SZ-X}\sigmalnzeta\sigmalnyx & \rho_\mathrm{WL-X}\sigmawl\sigmalnyx & \sigma_{\ln Y_\mathrm X}^2
    \end{pmatrix}
  \end{split}
\end{equation}
with correlation coefficients $\rho_\mathrm{SZ-X}$, $\rho_\mathrm{SZ-WL}$, and
$\rho_\mathrm{WL-X}$. With this, the full description of the
multi-observable--mass relation is
\begin{equation}
  \begin{split}
    P\Bigl(
    \begin{bmatrix}
      \ln\zeta \\ \ln M_\mathrm{WL} \\ \ln\yx
    \end{bmatrix} |&M,z,\vec p\Bigr) = \\
    &\mathcal N\Bigl(
    \begin{bmatrix}
      \langle\ln\zeta\rangle(M,z,\vec p) \\ \langle\ln M_\mathrm{WL}\rangle(M,z,\vec p) \\ \langle \ln\yx\rangle(M,z,\vec p)
    \end{bmatrix}
    , \vec\Sigma_\text{multi-obs}\Bigr).
  \end{split}
\end{equation}
All parameters of the observable--mass relations are listed in
Table~\ref{tab:parameters}.

While our default X-ray observable is \yx, we also consider the X-ray gas mass
\mgas. Note that both observables share the same \mgas\ data, and so we do not
use them simultaneously. We define a relation for the gas mass fraction
$\fgas\equiv \mgas/M_{500c}$
\begin{equation}
  \begin{split}
    \langle\ln\fgas\rangle =& \ln\left(\frac{\amg}{h_{70}^{3/2}}\right) + (\bmg-1) \ln\left(\frac{M_{500c} \, h_{70}}{5\times 10^{14}M_\odot}\right) \\
    &+ \cmg\ln\left(\frac{E(z)}{E(0.6)}\right)
  \end{split}
\end{equation}
with which the \mgas--mass relation becomes
\begin{equation}
\begin{split}
\langle\ln\left(\frac{\mgas}{5\times 10^{14}M_\odot}\right)\rangle =& \ln\left(\frac{\amg}{h_{70}^{5/2}}\right) + \bmg\ln\left(\frac{M_{500c} \, h_{70}}{5\times 10^{14}M_\odot}\right) \\
&+ \cmg\ln\left(\frac{E(z)}{E(0.6)}\right).
\label{eq:mgas}
\end{split}
\end{equation}

\subsubsection{The SZ $\xi$--mass Relation}

The observable we use to describe the cluster SZ signal is $\xi$, the detection
signal-to-noise ratio (SNR) maximized over all filter scales. To account for the
impact of noise bias, the \textit{unbiased} SZ significance $\zeta$ is
introduced, which is the SNR at the true, underlying cluster position and filter
scale \citep{vanderlinde10}. Following previous SPT work, 
$\xi$
across many noise realizations is related to $\zeta$ as
\begin{equation} \label{eq:xi2zeta}
P(\xi|\zeta) = \mathcal N(\sqrt{\zeta^2 + 3}, 1)
\end{equation}
In practice, we only map objects with $\zeta>2$ to $\xi$ using this relation,
but the exact location of this cut has no impact on our results \citepalias[see
also][]{dehaan16}. The validity of this approach and of Eq.~\ref{eq:xi2zeta} has
been extensively tested and confirmed by analyzing simulated SPT observations of
mock SZ maps \citep{vanderlinde10}.

The SPT-SZ survey consists of 19 fields that were observed to different depths.
The varying noise levels only affect the normalization of the $\zeta$--mass
relation, and leave \bsz, \csz, and \sigmalnzeta\ effectively unchanged
\citepalias{dehaan16}. In the analysis presented here, \asz\ is rescaled by a
correction factor for each of the 19 fields, which then allows us to work with a
single SZ observable--mass relation, given by Eq.~\ref{eq:zetaM}. The scaling
factors $\gamma_\mathrm{field}$ can be found in Table~1 in
\citetalias{dehaan16}.

In a departure from previous SPT analyses, we do not apply informative
(Gaussian) priors on the SZ scaling relation parameters. The self-calibration
through fitting the cluster sample against the halo mass function, \citep[see,
e.g.,][]{majumdar04}, the constraint on the normalization of the
observable--mass relations through our WL data, and the constraint on the SZ
scatter through the X-ray data are strong enough to constrain all four SZ
scaling relation parameters (in \nuLCDM, see Table~\ref{tab:constraints}). When
not including the X-ray data in our fit, however, we apply a Gaussian prior
$\sigmalnzeta=0.13\pm0.13$ as in \citetalias{dehaan16} \citep[this constraint
was extracted from mock observations of hydrodynamic simulations
from][]{lebrun14}.

We discuss possible limitations in our description of the $\xi$--mass relation
that would lead to systematic biases in the recovered cosmological constraints.
Because of our empirical weak-lensing mass calibration and the parametrization
of the SZ scaling relation by power laws and lognormal scatter with free
parameters, any bias in the SZ--mass relation that can be described by a power
law and/or lognormal scatter would only lead to parameter shifts in the SZ
scaling relation, but would not affect \emph{cosmological} parameter
constraints. Therefore, important systematics would be from potential
contaminants that would lead to an additional, non-lognormal scatter, a mass or
redshift dependence in the scatter, or a redshift dependence of the mass-slope.

A potential worry might be the dilution of the SZ signal by AGN activity and the
presence of dusty star-forming galaxies in the cluster. Various studies have
found that emission by dusty star-forming galaxies is negligible compared to the
SZ signal (see, e.g., \cite{lin09, sehgal10} and the summary in Section~6.4 in
\cite{benson13}). \cite{gupta17b} measured the cluster radio luminosity function
using an X-ray selected cluster sample at $z\lesssim0.7$ and concluded that
radio sources obeying this luminosity function would not have a strong impact on
the SZ signal. Only a few percent at most of the SPT-SZ clusters would host
sufficiently bright radio sources for their SZ signal to drop below the
selection threshold, and this is within the Poisson uncertainty of our sample.
At higher redshifts, it has been previously measured that the radio fraction in
optically selected clusters somewhat decreases at $z>0.65$ \citep{gralla11}.
This result is consistent with simulations of the microwave sky from
\cite{sehgal10}, which predicted that the amount of radio contamination in SZ
surveys was either flat or falling at $z>0.8$. Using tests against mocks, we
find for example that, to cause a shift in $w$ by more than $\Delta(w)=-0.3$,
the level of SZ contamination would have to be strong enough to remove more than
$\sim30\%$ of all cluster detections at redshifts $z\gtrsim1$, which by far
exceeds the measurement by \cite{gupta17b}. In conclusion, none of the discussed
sources of potential SZ cluster contamination have an impact that is strong
enough to introduce large biases in our cosmological constraints.

Another approach to testing the robustness of the SZ observable--mass relation
is to compare it with other cluster mass proxies, and to try and find deviations
from the simple scaling relation model. Note that, if such a deviation was
found, it would be hard to discern which observable is behaving in an unexpected
way, but importantly, one would learn that the multi-observable model needs an
extension. At low and intermediate redshifts $z\lesssim0.8$, comparisons with
cluster samples selected through optical and X-ray methods have shown that the
cluster populations can be described by power-law observable--mass scaling
relations with lognormal intrinsic scatter \citep{vikhlinin09b, mantz10b,
saro15, mantz16, saro17}. At higher redshifts, the subset of the SPT selected
sample with available X-ray observations from \chandra\ and {\it XMM-Newton}
exhibit scaling relations in X-ray $T_\mathrm{X}$, \yx, \mgas, and
$L_\mathrm{X}$ as well as in stellar mass galaxies, that are consistent with
power-law relations in mass and redshift with lognormal intrinsic scatter
\citep{chiu16,hennig17,chiu18,bulbul19}. When a redshift dependent mass slope
parameter has been included in the analyses of these datasets, the parameter
constraints have been statistically consistent with 0 in all cases \citep[see
Table 4 in][]{bulbul19}.

In conclusion, our description of the $\xi$--mass relation has been confirmed by
various independent techniques, especially for redshifts $z\lesssim1$. Note that
these tests are harder to perform at higher redshifts where non-SZ selected
samples are small and more challenging to characterize. Our expectation is that
as the cluster sample grows larger and the mass calibration information improves
that we will be able to characterize the currently negligible departures from
our scaling relation model. At that point, we will need to extend our
observable--mass relation to allow additional freedom.

\subsubsection{The Weak-Lensing Observable--Mass Relation}
\label{sec:MwlM}

The WL modeling framework used in this work is introduced in
\citetalias{dietrich19}, and we refer the reader to their Section~5.2 for
details.

The WL observable is the reduced tangential shear profile
$g_\mathrm{t}(\theta)$, which can be analytically modeled from the halo mass
$M_{200c}$, assuming an NFW halo profile and using the redshift distribution of
source galaxies \citep{wright00}. Miscentering, halo triaxiality, large-scale
structure along the line of sight, and uncertainties in the concentration--mass
relation, introduce bias and/or scatter. As introduced in Eq.~\ref{eq:MwlM}, we
assume a relation $\ln M_\mathrm{WL} = \ln(\bwl M_\mathrm{true})$, and use
numerical simulations to calibrate the normalization \bwl\ and the scatter about
the mean relation. Our WL dataset consists of two subsamples (Megacam and HST)
with different measurement and analysis schemes. We expect some systematics to
be shared among the entire sample, while others will affect each subsample
independently.

\begin{deluxetable*}{llcc}
\tablecaption{
\label{tab:WLmodeling}
WL modeling parameters \citepalias{schrabback18b, dietrich19}. The \emph{WL mass
bias} and the  (lognormal) \emph{intrinsic scatter} are calibrated against
$N$-body simulations. Among other effects, they also account for the uncertainty
and the scatter in the $c(M)$ relation. This is done separately for each cluster
in the HST sample leading to a range of values; here we report the smallest and
largest individual values. The \emph{mass modeling uncertainty} accounts for
uncertainties in the calibration against $N$-body simulations and in the
centering distribution. The \emph{systematic measurement uncertainties} account
for a multiplicative shear bias and the uncertainty in estimating the redshift
distribution of source galaxies. \emph{Uncorrelated large-scale structure} along
the line of sight leads to an additional, Gaussian scatter.}
\tablehead{
{Effect} & Parameter & \multicolumn{2}{c}{Impact on Mass} \\
& & Megacam & HST}
\startdata
Intrinsic scatter & $\sigma_\mathrm{intrinsic}$ & $0.214$ & $0.26-0.42$\\
$\Delta$(Intrinsic scatter) & $\Delta\sigma_\mathrm{intrinsic}$ & $0.04$ & $0.021-0.055$\\
Uncorrelated LSS scatter & $\sigma_\mathrm{LSS}$ & $9\times10^{13}M_\odot$ & $8\times10^{13}M_\odot$ \\
$\Delta$(Uncorrelated LSS scatter) & $\Delta\sigma_\mathrm{LSS}$ & $10^{13}M_\odot$ & $10^{13}M_\odot$ \\
\tableline
WL mass bias & $b_\text{WL mass}$ & $0.938$ & $0.81-0.92$ \\
Mass modeling uncertainty & $\Delta b_\text{WL mass model}$ & $4.4\%$ & $5.8-6.1\%$ \\
Systematic measurement uncertainty & $\Delta b_\text{measurement systematics}$ & $3.5\%$ & $7.2\%$ \\
Total systematic uncertainty & N/A & $5.6\%$ & $9.2-9.4\%$
\enddata
\end{deluxetable*}

We model the WL bias as
\begin{equation}
\begin{split}
b_{\mathrm{WL},i} &= b_{\text{WL mass},i} \\
&+ \delta_\mathrm{WL,bias} \, \Delta b_{\text{WL mass model},i} \\
&+ \delta_i \, \Delta b_{\text{measurement systematics},i}, \\
&i\in \{\text{Megacam, HST}\},
\end{split}
\end{equation}
where $b_\text{WL mass}$ is the mean bias due to WL mass modeling, $\Delta
b_\text{WL mass model}$ is the uncertainty on $b_\text{WL mass}$, and $\Delta
b_\text{measurement systematics}$ is the systematic measurement uncertainty due
to multiplicative shear bias and uncertainties in the determination of the
source redshift distribution; $\delta_\mathrm{WL,bias}$,
$\delta_\mathrm{Megacam}$, and $\delta_\mathrm{HST}$ are free parameters in our
likelihood. With this parametrization, we apply Gaussian priors $\mathcal
N(0,1)$ on the three fit parameters. The numerical values of the different
components of the WL bias are given in Table~\ref{tab:WLmodeling}.

The width of the (lognormal) scatter that is intrinsic to fitting WL shear
profiles against NFW profiles is
\begin{equation}
\begin{split}
\sigma_{\mathrm{WL},i} &= \sigma_{\mathrm{intrinsic},i} + \delta_\mathrm{WL,scatter} \, \Delta\sigma_{\mathrm{intrinsic},i}, \\
&i\in \{\text{Megacam, HST}\},
\end{split}
\end{equation}
where $\sigma_\mathrm{intrinsic}$ and $\Delta\sigma_\mathrm{intrinsic}$ are the
mean intrinsic scatter and the error on the mean (given in
Table~\ref{tab:WLmodeling}); $\delta_\mathrm{WL,scatter}$ is a free parameter in
our likelihood on which we apply a Gaussian prior $\mathcal N(0,1)$.

Finally, the width of the (normal) scatter due to uncorrelated large-scale
structure is
\begin{equation}
\begin{split}
\sigma_{\mathrm{WL,LSS},i} &= \sigma_{\mathrm{LSS},i} + \delta_{\mathrm{WL,LSS},i} \, \Delta\sigma_{\mathrm{LSS},i}, \\
&i\in \{\text{Megacam, HST}\},
\end{split}
\end{equation}
with the mean scatter $\sigma_\mathrm{LSS}$ and the error on the mean
$\Delta\sigma_\mathrm{LSS}$ given in Table~\ref{tab:WLmodeling} and where we
apply a Gaussian prior $\mathcal N(0,1)$ on the fit parameters
$\delta_{\mathrm{WL,LSS}_\mathrm{Megacam}}$ and
$\delta_{\mathrm{WL,LSS}_\mathrm{HST}}$.

For reference, the total systematic error in the WL calibration is $5.6\%$ for
the Megacam sample \citepalias{dietrich19} and $9.2-9.4\%$ for the HST sample
\citepalias{schrabback18b}. Given the small sample size of 19 and 13 clusters,
our WL mass calibration is still dominated by statistical errors.

\subsection{Likelihood Function}

The analysis pipeline used in this work evolved from the code originally used in
a previous SPT analysis \citep{bocquet15}. Since then, we have updated it to the
full 2500 deg$^2$ survey, included the handling of WL data and the
ability to account for correlated scatter among all observables, and modified
the X-ray analysis (see Section~\ref{sec:updateXrayAnalysis}). The pipeline is
written as a module for \textsc{CosmoSIS} \citep{zuntz15} and was also used for
other WL scaling relation studies of SPT-SZ clusters
\citep[\citetalias{dietrich19};][]{stern19}.

We start from a multi-observable Poisson log-likelihood
\begin{equation}\begin{split}\label{eq:likelihood_start}
\ln \mathcal L(\vec p) = & \sum_i \ln\frac{dN(\xi, \yx, g_\mathrm{t}, z |\vec p)}{d\xi d\yx dg_\mathrm{t} dz}
  \big|_{\xi_i, Y_{\mathrm{X}_i}, g_{\mathrm{t}_i}, z_i} \\
& - \iiiint d\xi\, d\yx\, dg_\mathrm{t}\, dz\, \left [ \right. \\
& \frac{dN(\xi, \yx, g_\mathrm{t}, z |\vec p)}{d\xi d\yx dg_\mathrm{t} dz} \Theta_\mathrm{s}  \left.\right ]\\
& +\mathrm{const.}
\end{split}\end{equation}
where the sum runs over all clusters $i$ in the sample, and
$\Theta_\mathrm{s}$ is the survey selection function; in our case
$\Theta_\mathrm{s}=\Theta(\xi>5, z>0.25)$.

As discussed in \cite{bocquet15} and explicitly shown in their Appendix, we
rewrite the first term in Eq.~\ref{eq:likelihood_start} as $P(\yx, g_\mathrm{t}
| \xi_i, z_i, \vec p)\big|_{Y_{\mathrm{X}_i},\, g_{\mathrm{t}_i}} \times
\frac{dN(\xi, z | \vec p)}{d\xi dz} \big|_{\xi_i, z_i}$. The second term in
Eq.~\ref{eq:likelihood_start} represents the total number of clusters in the
survey, which are selected in $\xi$ and $z$ (and without any selection based on
the follow-up observables). Therefore, this term reduces to $\int d\xi dz
\Theta_\mathrm{s} dN(\xi, z | \vec p)/d\xi dz$. With these modifications, and
after explicitly setting the survey selection, the likelihood function becomes
\begin{equation}
  \begin{split} \label{eq:likelihood}
    \ln \mathcal L(\vec p) =  & \sum_i \ln \frac{dN(\xi, z| \vec p)}{d\xi dz} \big|_{\xi_i, z_i} \\
    &- \int_{z_\mathrm{cut}}^\infty dz \int_{\xi_\mathrm{cut}}^\infty d\xi \frac{dN(\xi, z| \vec p)}{d\xi dz} \\
    &+ \sum_j \ln P(\yx, g_\mathrm{t} | \xi_j, z_j, \vec p)\big|_{Y_{\mathrm{X}_j},\, g_{\mathrm{t}_j}}
  \end{split}
\end{equation}
up to a constant. The first sum runs over all clusters $i$ in the sample, and
the second sum runs over all clusters $j$ with \yx\ and/or WL $g_\mathrm{t}$
measurements.

\begin{figure*}
\includegraphics[width=\textwidth]{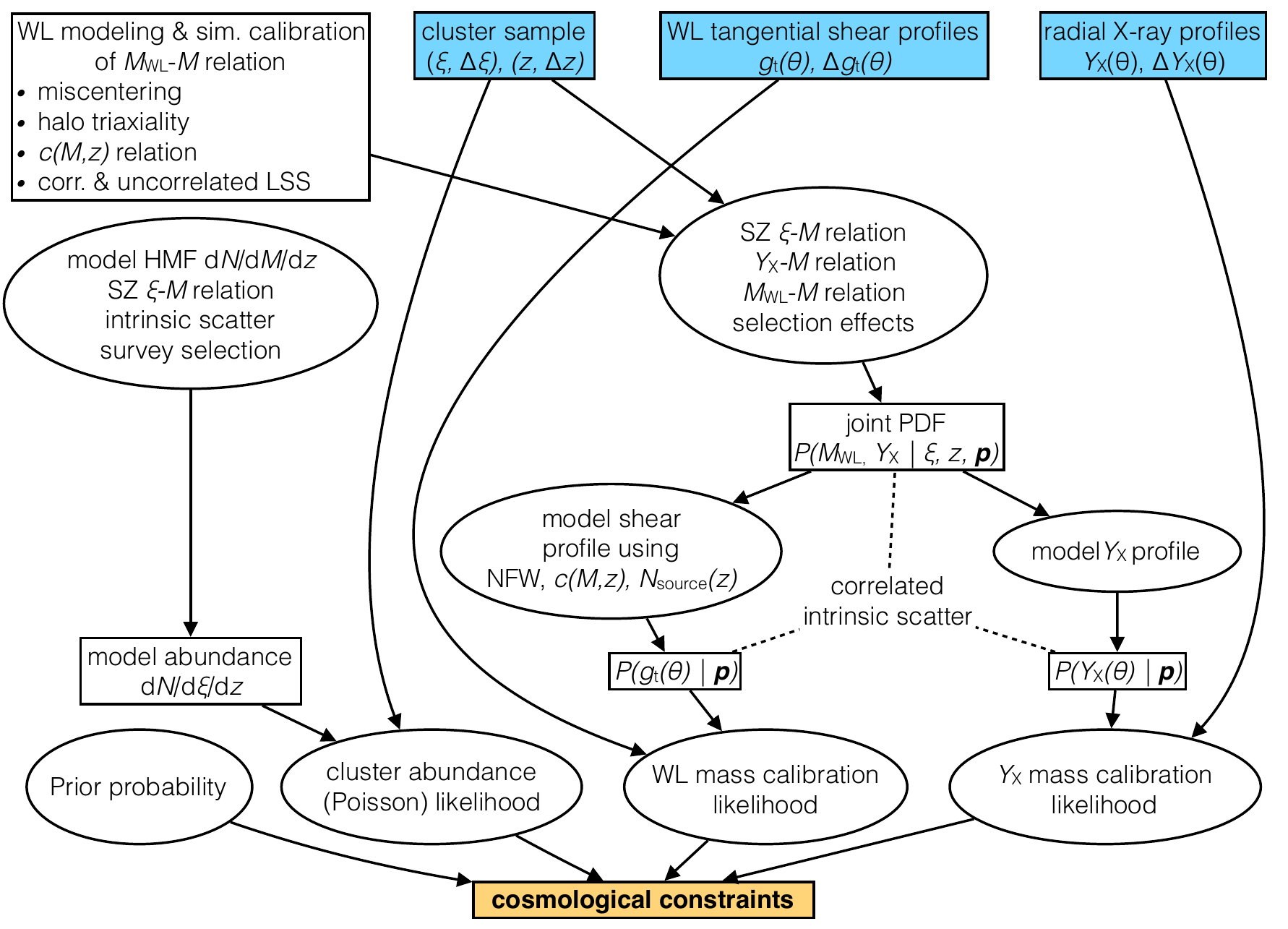}
\caption{Analysis flowchart showing how the cluster data (blue boxes) are used
to obtain cosmological constraints (orange box). White boxes show model
predictions, ellipses show functions that use or create those models. The number
count analysis is performed using the full SPT-SZ catalog, while the mass
calibration is performed using the subset of clusters for which follow-up data
is available.}
\label{fig:pipeline}
\end{figure*}

The first two terms in Eq.~\ref{eq:likelihood} can be interpreted as the
likelihood of the abundance (or number counts) of SZ clusters, while the third
term represents the information from follow-up mass calibration. These two
components are also visualized in the analysis flowchart in
Fig.~\ref{fig:pipeline}: the number counts on the lower left side use the
distribution of clusters in $(\xi,z)$ space, and the mass calibration on the
lower right also uses all available WL and X-ray follow-up data.

We note that the subsamples of clusters that were targeted for follow-up WL
and/or X-ray  data were selected at random within some cuts in $\xi$ and
redshift. Importantly, the selection was not made on WL and/or X-ray
measurements. Therefore, the likelihood function presented above is complete;
importantly, it does not suffer  from biases from WL and/or X-ray selections.

\subsubsection{Implementation of the Likelihood Function}
\label{sec:likelihood_implement}

We compute the individual terms in Eq.~\ref{eq:likelihood} as follows.
\begin{equation} \label{eq:dN_dxidz}
\begin{split}
\frac{dN(\xi, z |\, \vec p)}{d\xi dz} =
\iint & dM\, d\zeta\, \left [ \right. P(\xi|\zeta)P(\zeta|M,z,\vec p)\\
&\frac{dN(M, z| \vec p)}{dM dz}\Omega(z,\vec p) \left.\right ]
\end{split}
\end{equation}
where $\Omega(z,\vec p)$ is the survey volume and $dN(M, z| \vec p)/dM dz$ is
the HMF. We evaluate Eq.~\ref{eq:dN_dxidz} in the space $(\xi,z)$ by
convolving the HMF with the intrinsic scatter in $P(\zeta|M,z,\vec p)$ and the
measurement uncertainty in $P(\xi|\zeta)$.

The first term in Eq.~\ref{eq:likelihood} is computed by evaluating
Eq.~\ref{eq:dN_dxidz} at each cluster's measured $(\xi_i,z_i)$, marginalizing
over photometric redshift errors where present. The second term is a simple
two-dimensional integral over Eq.~\ref{eq:dN_dxidz}.

Our cluster sample contains 22 SZ detections for which no optical counterparts
were found; these were assigned lower redshift limits $z_\mathrm{lim}$ in
\cite{bleem15b}. We used simulations to determine the expected false-detection
rate $dN_\text{false}(\xi)/d\xi$ given survey specifics \citepalias[see
Section~2.2 and Table~1 in][]{dehaan16}. For each unconfirmed cluster
candidates, we evaluate a modified version of the first term in
Eq.~\ref{eq:likelihood}
\begin{equation}
\begin{split}
\frac{dN_\text{unconf. cand.}(\xi, z| \vec p)}{d\xi dz} = & \frac{dN_\text{cluster}(\xi, z| \vec p)}{d\xi dz} \\
  & + \frac{dN_\text{false}(\xi)}{d\xi}
\end{split}
\end{equation}
and marginalize over the candidate's allowed redshift range
$z_\mathrm{lim}<z<\infty$. Note that the total expected number of false
detections $\int d\xi dN_\text{false}(\xi)/d\xi$ is independent of $\vec p$ and
is therefore neglected in Eq.~\ref{eq:likelihood}. The expected number of false
detections in the SPT-SZ survey is $18\pm4$, which is consistent with our 22
unconfirmed candidates \citepalias{dehaan16}. In practice, we obtain essentially
unchanged results if we simply discard the 22 optically-unconfirmed SZ
detections from the catalog. There are nine clusters that are detected in the
overlap region between adjacent SPT fields. We follow  \citetalias{dehaan16} and
double-count these clusters in our analysis. Accounting for only one object of
each pair of these clusters instead does not change our results in any
significant way.

The mass calibration term in Eq.~\ref{eq:likelihood} is computed as
\begin{equation}
\label{eq:masscalib}
\begin{split}
 P(\yx^\mathrm{obs}, & g_\mathrm{t}^\mathrm{obs} | \xi, z, \vec p) = \\
  &  \iiiint dM\, d\zeta\, d\yx\, dM_\mathrm{WL}\, \left [ \right. \\
 &P(\yx^\mathrm{obs}|\yx)
  P(g_\mathrm{t}^\mathrm{obs}|M_\mathrm{WL})
 P(\xi|\zeta) \\
 & P(\zeta,\yx,M_\mathrm{WL}|M,z,\vec p) P(M|z,\vec p)
   \left. \right ]
\end{split}
\end{equation}
with the HMF $P(M|z,\vec p)$ and the multi-observable scaling relation
$P(\zeta,\yx,\mwl|M,z,\vec p)$ that includes the effects of correlated scatter.
Computing this multi-dimensional integral in the $(\zeta,\yx,\mwl)$ space is
expensive. We minimize the computational cost of this step by i) only
considering parts of the $(\zeta,\yx,\mwl)$ space that have non-negligible
probability densities; we estimate this sub-space from the measurements and
$\vec p$, ii) using Fast Fourier Transform convolutions, and iii) only
performing this computation for clusters that actually have both follow-up
measurements \yx\ and \mwl; otherwise, we restrict the computation to the much
cheaper two-dimensional $(\yx,\zeta)$ or $(\mwl,\zeta)$ spaces. The
mass calibration term does not need to be computed at all for clusters that have
no X-ray or WL follow-up data.

\subsubsection{Update of the X-ray Analysis Scheme}
\label{sec:updateXrayAnalysis}

The X-ray observable is a measurement of the radial \yx\ profile. The scaling
relation on the other hand predicts a value of the observable integrated out to
$r_{500}$ for a given $M_{500}$. In a self-consistent analysis, the likelihood
should be extracted by comparing the data and the model prediction at the same
radius.

In previous SPT analyses, a \yx\ value was extracted from the profile by
iteratively solving for the radius $r_\mathrm{iter}$ at which the measured \yx\
and the X-ray scaling relation prediction from Eq.~\ref{eq:YxM} match (the
scaling relation is evaluated at $M_{500}\equiv4\pi/3r_\mathrm{iter}^3 500
\rho_c$). This iteration was repeated for each set of parameters $\vec p$, but
within a fixed reference cosmology. However, this method introduces a bias,
because $r_\mathrm{iter}$ is not equal to the radius $r_{500}$ at which the
scaling relation $P(\zeta,\yx,M_\mathrm{WL}|M_{500},z,\vec p)$ in
Eq.~\ref{eq:masscalib} is evaluated.

We choose a different approach, and evaluate both the (integrated) measured
profile and the model prediction at a fixed fiducial radius $r_\mathrm{fid}$. We
define $r_\mathrm{fid}$ for each cluster by computing $r_{500,\mathrm{fid}}$
from its SZ significance $\xi$ using a fiducial set of SZ scaling relation
parameters, and setting $r_\mathrm{fid}=r_{500,\mathrm{fid}}$. Then, for each
set of parameters $\vec p$ in the analysis, we convert the model prediction
$\yx(r_{500})$ from radius $r_{500}$ to $r_\mathrm{fid}$. We use the fact that
the radial profiles are well-approximated by power laws in radius
\begin{equation}
\frac{\yx(r)}{\yx(r_{500})} = \left(\frac{r}{r_{500}}\right)^{d\ln \yx/d\ln r}
\end{equation}
where $r_{500}$ is derived from $M_{500c}$. In our analysis, we assume
isothermality (see Section~\ref{sec:Xraydata}), and so $d\ln \yx/d\ln r$ equals
the radial slope in gas mass $d\ln M_\mathrm{g}/d\ln r$. From our sample we
measure
\begin{equation}
d\ln M_\mathrm{g}/d\ln r = 1.12 \pm 0.23.
\end{equation}
We are now able to make a model prediction at $r_\mathrm{fid}$, starting from
the scaling relation prediction $\yx(r_{500})$:
\begin{equation}
\yx(r_\mathrm{fid}) = \yx(r_{500})
  \left(\frac{r_\mathrm{fid}}{r_{500}}\right)^{d\ln M_\mathrm{g}/d\ln r}.
\end{equation}
In the analysis, we marginalize over the uncertainty in $d\ln M_\mathrm{g}/d\ln
r $, which shows negligible correlation with any other parameter. Note that this
prescription for the model prediction $\yx(r_\mathrm{fid})$ contains an
additional dependence on $r_{500}$ and thus on $M_{500}$.

We note that a similar approach was adopted by other groups
\citep[e.g.][]{mantz10b, mantz15}. We have shown through tests against mock
catalogs that the new analysis scheme is unbiased, and that the previous method
biased \byx\ low at a level that is comparable to the uncertainty on that
parameter, while the effect on other parameters was very small.

\subsection{The Halo Mass Function}
\label{sec:HMF}

We assume the HMF fit by \cite{tinker08}. This approach assumes universality of
the HMF across the cosmological parameter space considered in this work, and
uses a fitting function that was calibrated against $N$-body simulations. In
principle, the HMF is also affected by baryonic effects. However, hydrodynamic
simulations suggest that these have negligible impact for clusters with masses
as high as those considered here \citep{velliscig14}; this was explicitly tested
for a simulated and idealized SPT-SZ cluster survey \citep{bocquet16a}. Finally,
note that the \cite{tinker08} fit applies to \emph{mean} spherical overdensities
in the range $200\leq\Delta_\mathrm{mean}\leq3200$, and we thus convert to
$\Delta_{500\mathrm{crit}}$ using $\Delta_\mathrm{mean}(z) = 500/\Om(z)$. As the
HMF fit is only calibrated up to $\Delta_\mathrm{mean}=3200$, we require
$\Om(z)\geq500/3200=0.15625$ for all redshifts $z\geq0.25$ relevant for our
cluster sample.

\subsection{Pipeline Validation on Mock Data}
\label{sec:validation}

We have run extensive tests to ensure that our analysis pipeline is unbiased at
a level that is much smaller than our total error budget. The primary approach
is testing against mock catalogs. Of course such tests are only useful if
producing mocks is easier and more reliable than the actual analysis. In our
case, the analysis is challenging mainly because of the computation of
multi-dimensional integrals. To create one of our mocks on the other hand, one
has to compute the halo mass function, apply the observable--mass relations,
draw random deviates, and compute WL shear profiles. Using the same code to
compute the HMF for the mocks and the analysis would undercut the usefulness of
the testing, and so we also created mocks using HMFs computed with independent
code. For the same reason, the mock shear profiles were created using
independent code. We typically create mock catalogs that contain an order of
magnitude more clusters and calibration data than our real sample. We created
and analyzed sets of mocks using different random seeds and different sets of
input parameters (notably, some with $w\neq-1$). No test indicated any biases in
our analysis pipeline at the level relevant for our data set.

\subsection{Quantifying Posterior Distribution (Dis-)Agreement}
\label{sec:quantify_agreement}

We characterize the agreement between constraints obtained from pairs of probes
(e.g., clusters and primary CMB anisotropies) by quantifying whether the
difference between the two posterior distributions is consistent with zero
difference. We draw representative samples $[\vec x_1]$ and $[\vec x_2]$ from
the posteriors of the two probes $P_1(\vec x)$ and $P_2(\vec x)$, compute the
difference between all pairs of points $\vec \delta \equiv \vec x_1 -\vec x_2$
and then construct the probability distribution $D$ from the ensemble $[\vec
\delta]$. The probability value (or $p$-value) that the two distributions
represent the same underlying quantity is
\begin{equation}
p = \int\limits_{D < D(0)} d\vec \delta D(\vec \delta)
\end{equation}
where $D(\vec 0)$ is the probability of zero difference. The $p$-value can be
converted into a significance assuming Gaussian statistics. This measure can be
applied to one-dimensional and multi-dimensional parameter spaces. The code is
publicly
available.\footnote{\url{https://github.com/SebastianBocquet/PosteriorAgreement}}

\subsection{Parameter Priors and Likelihood Sampling}

In our cosmological fits, we assume spatial flatness and allow the sum of
neutrino masses to vary. The comparison of our results with constraints from
primary CMB anisotropies is of prime interest---notably, the comparison of
constraints on \sig. For primary CMB anisotropies, \sig\ is strongly degenerate
with \sumMnu\ and so the latter should be a free parameter of the model to avoid
artificially tight constraints. We refer to the flat \LCDM\ model with a varying
sum of neutrino masses as \nuLCDM, and to its extension with a free dark energy
equation of state parameter as \nuwCDM.

In the \nuLCDM\ cosmology, we vary the cosmological parameters \Om, \Omnuhh,
\Ombhh, $A_s$, $h$, $n_s$; \sig\ is a derived parameter. Our cluster data
primarily constrain \Om\ and \sig, and we marginalize over flat priors on the
other parameters. The parameter ranges for \Ombhh\ and $n_s$ are chosen to
roughly match the $5\sigma$ credibility interval of the \planck\ constraints;
$h$ is allowed to vary in the range $0.55\dots0.9$. We assume two massless and
one massive neutrino and allow \Omnuhh\ to vary in the range $0\dots 0.01$; this
corresponds to a range in \sumMnu\ of $0\dots0.93$~eV. We note that the minimum
allowed sum of neutrino masses from oscillation experiments is
$\sumMnu>59.5\pm0.5$~meV \citep{tanabashi18}. In a departure from previous SPT
analyses, we do not apply a BBN prior on \Ombhh\ or constraints from direct
measurements of $H_0$. We remind the reader that the implementation of the
theory HMF leads to an effective, hard prior $\Om(z)\gtrsim0.16$ for all
redshifts $z>0.25$ relevant to our survey (see Section~\ref{sec:HMF}); however,
this prior does not affect our results. All parameters and their priors are
summarized in Table~\ref{tab:parameters}.

The likelihood sampling is done within \textsc{CosmoSIS} using the
\textsc{Metropolis} \citep{metropolis53} and \textsc{MultiNest} \citep{feroz09}
samplers. We confirmed that they produce consistent results.

\begin{deluxetable}{ll}
\tablecaption{\label{tab:parameters}
Summary of cosmological and astrophysical parameters used in our fiducial
analysis. The Gaussian prior on \sigmalnzeta\ is only applied when no X-ray data
is included in the fit. The parameter ranges for \Ombhh\ and $n_s$ are chosen to
roughly match the $5\sigma$ interval of the \planck\ \LCDM\ results. $w$ is
fixed to $-1$ for \LCDM, and is allowed to vary for \wCDM. The optical depth to
reionization $\tau$ is only relevant when \planck\ data is included in the
analysis. The WL modeling systematics are presented in
Table~\ref{tab:WLmodeling}.}
\tablehead{\colhead{Parameter} & \colhead{Prior}}
\startdata
\multicolumn{2}{l}{Cosmological}\\
\Om & $\mathcal U(0.05,0.6),\, \Om(z>0.25)>0.156$ \\
\Ombhh\ & $\mathcal U(0.020,0.024)$ \\
\Omnuhh & $\mathcal U(0,0.01)$ \\
$\Omega_k$ & fixed ($0$) \\
$A_s$ & $\mathcal U(10^{-10}, 10^{-8})$ \\
$h$ & $\mathcal U(0.55,0.9)$ \\
$n_s$ & $\mathcal U(0.94,1.00)$ \\
$w$ & fixed ($-1$) or $\mathcal U(-2.5,-0.33)$\\
\tableline
\multicolumn{2}{l}{Optical depth to reionization}\\
$\tau$ & fixed or $\mathcal U(0.02,0.14)$ \\
\tableline
\multicolumn{2}{l}{SZ scaling relation}\\
\asz & $\mathcal U(1,10)$ \\
\bsz & $\mathcal U(1,2.5)$ \\
\csz & $\mathcal U(-1,2)$ \\
\sigmalnzeta & $\mathcal U(0.01,0.5)$ ($\times\mathcal N(0.13, 0.13^2))$\\
\tableline
\multicolumn{2}{l}{X-ray \yx\ scaling relation}\\
\ayx & $\mathcal U(3, 10)$ \\
\byx & $\mathcal U(0.3,0.9)$ \\
\cyx & $\mathcal U(-1,0.5)$ \\
\sigmalnyx & $\mathcal U(0.01,0.5)$\\
$d\ln M_\mathrm{g}/d\ln r$ & $\mathcal{U}(0.4, 1.8)\times \mathcal{N}(1.12, 0.23^2)$\\
\tableline
\multicolumn{2}{l}{WL modeling}\\
$\delta_\mathrm{WL,bias}$ & $\mathcal U(-3,3) \times \mathcal{N}(0,1)$ \\
$\delta_\mathrm{Megacam}$ & $\mathcal U(-3,3) \times \mathcal{N}(0,1)$ \\
$\delta_\mathrm{HST}$ & $\mathcal U(-3,3) \times \mathcal{N}(0,1)$ \\
$\delta_\mathrm{WL,scatter}$ & $\mathcal U(-3,3) \times \mathcal{N}(0,1)$ \\
$\delta_{\mathrm{WL,LSS}_\mathrm{Megacam}}$ & $\mathcal U(-3,3) \times \mathcal{N}(0,1)$ \\
$\delta_{\mathrm{WL,LSS}_\mathrm{HST}}$ & $\mathcal U(-3,3) \times \mathcal{N}(0,1)$ \\
\tableline
\multicolumn{2}{l}{Correlated scatter}\\
$\rho_\mathrm{SZ-WL}$ & $\mathcal U(-1,1)$ \\
$\rho_\mathrm{SZ-X}$ & $\mathcal U(-1,1)$ \\
$\rho_\mathrm{X-WL}$ & $\mathcal U(-1,1)$ \\
&$\det (\vec\Sigma_\text{multi-obs}) >0$
\enddata
\end{deluxetable}


\begin{deluxetable*}{l|ccc|ccc}
\tablecaption{\label{tab:constraints}
Constraints on a subset of cosmological and scaling relation parameters. SPTcl
stands for the SPT-SZ+WL+\yx\ dataset, and \planck\ refers to the TT+lowTEB
data. The cluster-based posterior distributions for $h$ and \sumMnu\ are poorly
constrained and strongly affected by the hard priors applied and we therefore do
not quote constraints.}
\tablehead{\colhead{Parameter} & \multicolumn{3}{c}{\nuLCDM} & \multicolumn{3}{c}{\nuwCDM}}
\startdata
 & SPT-SZ+WL & SPTcl & \planck+SPTcl & SPTcl & \planck+SPTcl & \planck+BAO+SNIa+SPTcl\\
 \Om & $0.285\pm0.047$ & $0.276\pm0.047$ & $0.353\pm0.027$ & $0.299\pm0.049$ & $0.347\pm0.039$ & $0.305\pm0.008$ \\
 \sig & $0.763\pm0.037$ & $0.781\pm0.037$ & $0.761\pm0.033$ & $0.766\pm0.036$ & $0.761\pm0.027$ & $0.801\pm0.026$ \\
 \sigOmtwo & $0.753\pm0.025$ & $0.766\pm0.025$ & $0.786\pm0.025$ & $0.763\pm0.024$ & $0.782\pm0.018$ & $0.803\pm0.024$ \\
 \sigOmfive & $0.739\pm0.041$ & $0.745\pm0.042$ & $0.824\pm0.020$ & $0.760\pm0.043$ & $0.816\pm0.032$ & $0.807\pm0.023$ \\
 $h$ & \nodata & \nodata & $0.645\pm0.019$ & \nodata & $0.657\pm0.039$ & $0.681\pm0.009$ \\
 \sumMnu\ [eV] & \nodata & \nodata & $0.39\pm0.19$ & \nodata & $0.50\pm0.24$& $0.16\pm0.10$ \\
 $w$ & $-1$ & $-1$ & $-1$ & $-1.55\pm0.41$ & $-1.12\pm0.21$ & $-1.03\pm0.04$ \\ [6pt]
 \asz & $5.68^{+0.89}_{-1.03}$ & $5.24^{+0.76}_{-0.93}$ & $4.58^{+0.63}_{-0.92}$ & $4.84^{+0.80}_{-0.97}$ & $4.57^{+0.55}_{-0.62}$ & $4.07^{+0.62}_{-0.76}$ \\
 \bsz & $1.519^{+0.087}_{-0.110}$ & $1.534^{+0.099}_{-0.100}$ & $1.667^{+0.069}_{-0.072}$ & $1.601^{+0.098}_{-0.102}$ & $1.653^{+0.079}_{-0.081}$ & $1.685^{+0.074}_{-0.088}$ \\
 \csz & $0.547^{+0.468}_{-0.375}$ & $0.465^{+0.492}_{-0.321}$ & $0.993^{+0.222}_{-0.218}$ & $1.290^{+0.443}_{-0.250}$ & $1.117^{+0.221}_{-0.191}$ & $0.746^{+0.165}_{-0.169}$ \\
 \sigmalnzeta & $0.152^{+0.066}_{-0.099}$ & $0.161^{+0.084}_{-0.075}$ & $0.162^{+0.083}_{-0.100}$ & $0.169^{+0.082}_{-0.072}$ & $0.148^{+0.073}_{-0.106}$ & $0.133^{+0.055}_{-0.133}$ \\
 \ayx & \dots & $6.35^{+0.68}_{-0.69}$ & $7.55^{+0.57}_{-0.56}$ & $6.33^{+0.69}_{-0.78}$ & $7.44^{+0.60}_{-0.68}$ & $7.38^{+0.63}_{-0.65}$ \\
 \byx & \dots & $0.514^{+0.032}_{-0.042}$ & $0.480^{+0.028}_{-0.035}$ & $0.499^{+0.032}_{-0.039}$ & $0.488^{+0.032}_{-0.037}$ & $0.480^{+0.033}_{-0.041}$ \\
 \cyx & \dots & $-0.310^{+0.140}_{-0.209}$ & $-0.464^{+0.131}_{-0.133}$ & $-0.669^{+0.120}_{-0.213}$ & $-0.525^{+0.141}_{-0.143}$ & $-0.371^{+0.123}_{-0.120}$ \\
 \sigmalnyx & \dots & $0.184^{+0.087}_{-0.089}$ & $0.180^{+0.095}_{-0.102}$ & $0.170^{+0.076}_{-0.094}$ & $0.205^{+0.094}_{-0.087}$ & $0.181^{+0.102}_{-0.162}$ \\
\enddata
\end{deluxetable*}

\section{Results}

Our fiducial results are obtained from the SPT-selected clusters with their
detection significances and redshifts, together with the WL and X-ray follow-up
data where available. We refer to this dataset as SPTcl (SPT-SZ+WL+\yx).

Constraints on cosmological and scaling relation parameters are summarized in
Table~\ref{tab:constraints}. We also provide constraints on the parameter
combination \sigOmtwo\ and \sigOmfive; the exponent $\alpha=0.2$ is chosen as it
minimizes the fractional uncertainty on $\sig (\Om/0.3)^\alpha$, and
$\alpha=0.5$ is common in other low-redshift cosmological probes.

\subsection{\nuLCDM\ Cosmology}

From the cluster abundance measurement of our SPTcl (SPT-SZ +WL+\yx) dataset we
obtain our baseline results
\begin{align}
\Om &= \LCDMclOm \\
\sig &= \LCDMclsig \\
\sigOmtwo &= \LCDMclsigOmtwo.
\end{align}
The remaining cosmological parameters (including \sumMnu, see
Fig.~\ref{fig:LCDM_compare_Planck}) are not or only weakly constrained by the
cluster data. Constraints on scaling relation parameters can be found in
Table~\ref{tab:constraints}. We note that applying priors on \Ombhh\ and $H_0$
from BBN and direct measurements of $H_0$ and/or fixing the sum of neutrino
masses to $0.06$~eV, approximately the lower limit predicted from terrestrial
oscillation experiments, does not affect our constraints on \Om\ and \sig\ in
any significant way (see Fig.~\ref{fig:LCDM_fix_nu} in the Appendix for the
impact of fixing the sum of the neutrino masses).

\subsubsection{Goodness of Fit}
\label{sec:goodnessoffit}

\begin{figure*}
  \includegraphics[width=\textwidth]{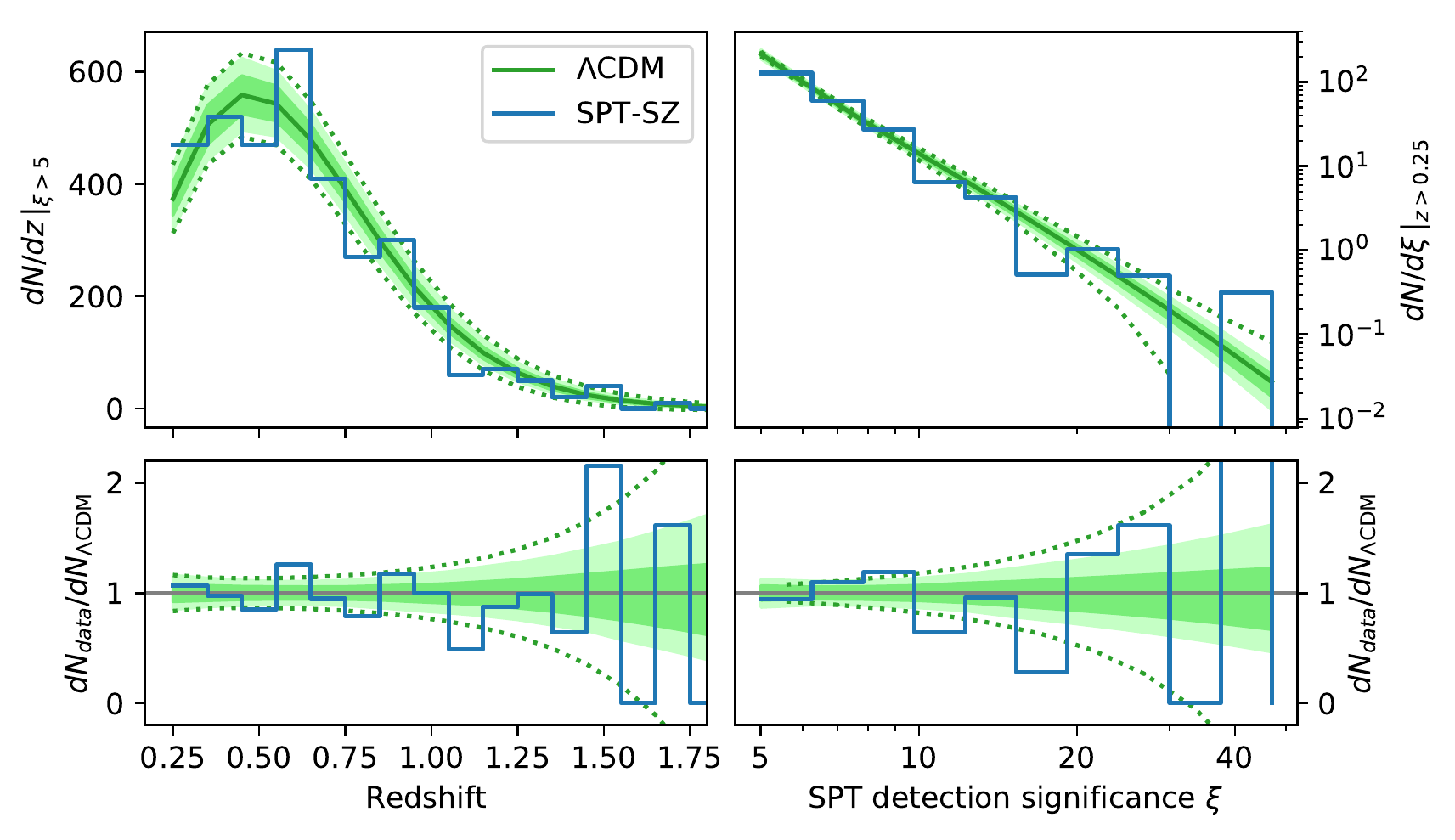}
  \caption{Distribution of clusters as a function of redshift (\emph{left
  panels}) and detection significance $\xi$ (\emph{right panels}). The \emph{top
  panels} show the SPT-SZ data and the recovered model predictions for \nuLCDM.
  The \emph{bottom panels} show the residuals of the data with respect to the
  model prediction. The different lines and shadings correspond to the mean
  recovered model and the $1\sigma$ and $2\sigma$ allowed ranges. The dotted
  lines show the Poisson error on the mean model prediction. There are no clear
  outliers and we conclude that the model provides an adequate fit to the data.}
  \label{fig:GoF}
\end{figure*}

In Fig.~\ref{fig:GoF}, we compare the measured distribution of clusters as a
function of their redshift and SPT detection significance with the model
prediction evaluated for the recovered parameter constraints. This figure does
not suggest any problematic feature in the data.

For a more quantitative discussion, we bin our confirmed clusters into a grid of
$30\times30$ in redshift and detection significance, and confront this
measurement with the expected number of objects in each two-dimensional bin. The
expected (and measured) numbers in each bin are too small to apply Gaussian
$\chi^2$ statistics, and we estimate the goodness of fit using a prescription
for the Poisson statistic \citep{kaastra17}.\footnote{We use the \textsc{python}
implementation from \url{https://github.com/abmantz/cstat}.} This approach is
similar to our likelihood analysis, which applies Poisson statistics within
infinitesimally small bins, instead of the larger bins we assume here. Adopting
the maximum-posterior \nuLCDM\ parameters, we compute the expected number of
clusters in each of the $30\times30$ bins and follow \cite{kaastra17} to
evaluate the test statistic $C$. We obtain an expected mean $C_\mathrm{e}$ and
variance
$C_\mathrm{v}$
\begin{equation}
C_\mathrm{e} = \LCDMCe; \,\, C_\mathrm{v} = \LCDMCv^2.
\end{equation}
For samples that contain at least a few hundred objects---like ours---the
statistic $C$ is well approximated by a Gaussian with mean $C_\mathrm{e}$ and
variance $C_\mathrm{v}$ \citep{kaastra17}. The data statistic for our sample is
\begin{equation}
C_\mathrm{d} = \LCDMCd
\end{equation}
in full agreement with the range expected for $C_\mathrm{e}$, indicating that
the model provides an adequate fit to the data.

\subsubsection{Comparison with Previous SPT results}

\begin{figure}
\includegraphics[width=\columnwidth]{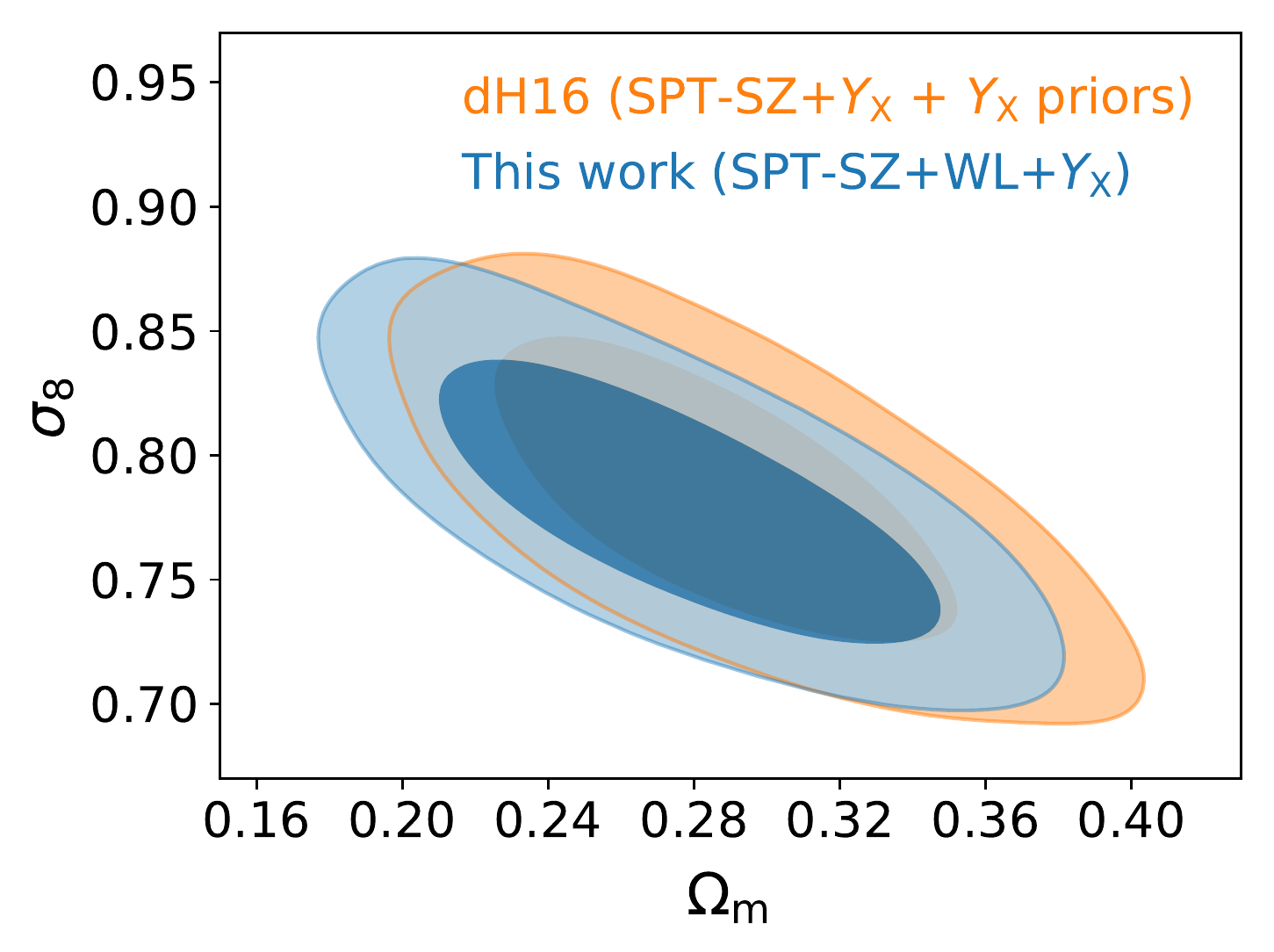}
\caption{Constraints on \Om\ and \sig\ from this analysis an from a previous
analysis that used the same cluster sample \citepalias{dehaan16}. The
consistency ($\dHsigma\sigma$) indicates that our internal mass calibration
using WL data agrees with the external X-ray mass calibration priors adopted in
\citetalias{dehaan16}.}
\label{fig:LCDM_compare_dH}
\end{figure}

As discussed in the Introduction, this work uses the same SPT-SZ cluster sample
\citep[][now with updated photometric redshifts, see
Section~\ref{sec:SPTsample}] {bleem15b} that was analyzed in
\citetalias{dehaan16}, and the key update is the inclusion of WL data. In
\citetalias{dehaan16}, the amplitude of the observable--mass relation was set by
a prior on the X-ray normalization \ayx, which in turn was informed by external
WL datasets \citep[CCCP and WtG,][]{applegate14, vonderlinden14a, hoekstra15}.
Gaussian priors were applied to the remaining SZ and X-ray scaling relation
parameters, which we dropped for this analysis. In
Fig.~\ref{fig:LCDM_compare_dH}, we compare our constraints on \Om-\sig\ with the
ones presented in \citetalias{dehaan16}. We recover very similar results; in
\Om-\sig\ space, the agreement is $p=\dHpvalue$ ($\dHsigma\sigma$). Since the
key difference between \citetalias{dehaan16} and this work is the inclusion of
WL data, this agreement indirectly confirms that our internal WL mass
calibration agrees with the external priors adopted previously. This is expected
because the X-ray prior adopted in previous work agrees well with the
measurement enabled by our own WL dataset \citepalias{dietrich19}.

\subsubsection{Comparison with External Probes}
\label{sec:externalprobes}

\begin{figure}
  \includegraphics[width=\columnwidth]{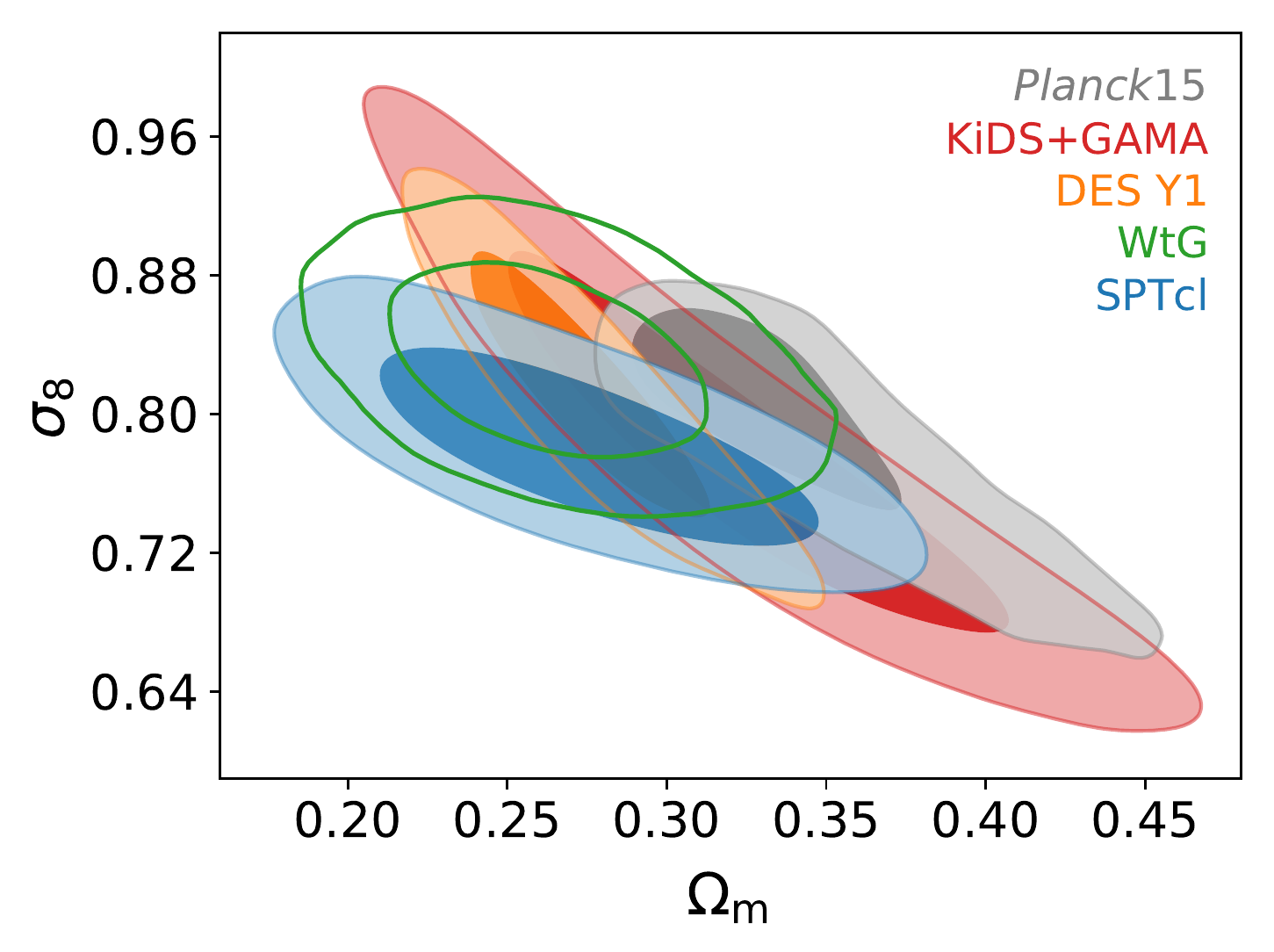}
  \caption{\nuLCDM\ constraints on \Om\ and \sig. The
  SPTcl dataset comprises SPT-SZ+WL+\yx, $Planck$ is TT+lowTEB, KiDS+GAMA and
  DES Y1 are cosmic shear+galaxy clustering+galaxy-galaxy-lensing. The WtG
  (X-ray selected clusters) result also contains their $f_\mathrm{gas}$
  measurement.}
  \label{fig:LCDM_compare_ext}
\end{figure}

In Fig.~\ref{fig:LCDM_compare_ext}, we show a comparison of our results with
constraints from \planck\ (TT+lowTEB) and from combined analyses of cosmic
shear, galaxy-galaxy lensing, and galaxy clustering from the Kilo Degree Survey
and the Galaxies And Mass Assembly survey \citep[KiDS+GAMA,][]{vanuitert18} and
the Dark Energy Survey (DES) Year 1 results \citep{des18-main}. We also compare
our results with another cluster study that used internal WL mass calibration,
but a sample based on X-ray selection \citep[Weighing the Giants, or
WtG,][]{mantz15}. Overall, the constraining power of all probes is roughly
similar in this plane. There is good agreement among all probes as the 68\%
contours all overlap. In particular, the cluster-based constraints yield very
similar \Om, but WtG favor a somewhat higher \sig. Interestingly, the degeneracy
axis of WtG is slightly tilted with respect to SPTcl, which we attribute to the
different redshift and mass ranges spanned by the two samples.

We pay particular attention to a comparison with \planck\ (TT+lowTEB). Our
constraint on $\sigOmtwo=\LCDMclsigOmtwo$ is lower than the one from \planck\
($\sigOmtwo=\LCDMPlancksigOmtwo$); the agreement between the two measurements is
$p=\nuLCDMclPlsigOmtwopvalue$ ($\nuLCDMclPlsigOmtwosigma\sigma$). In the
two-dimensional \Om-\sig\ space, the agreement is $p=\nuLCDMclPlpvalue$
($\nuLCDMclPlsigma\sigma$).

We note that the latest analysis of the cluster sample selected by the \planck\
satellite is qualitatively in agreement with our constraint, as shown in Fig.~32
in \cite{planck18-1}. Notably, the 95\% contour of their result, calibrated
using CMB lensing, encompasses the \planck\ primary CMB result in the \Om-\sig\
plane.

\subsubsection{Impact of X-ray Follow-up Data}

\begin{figure}
  \includegraphics[width=\columnwidth]{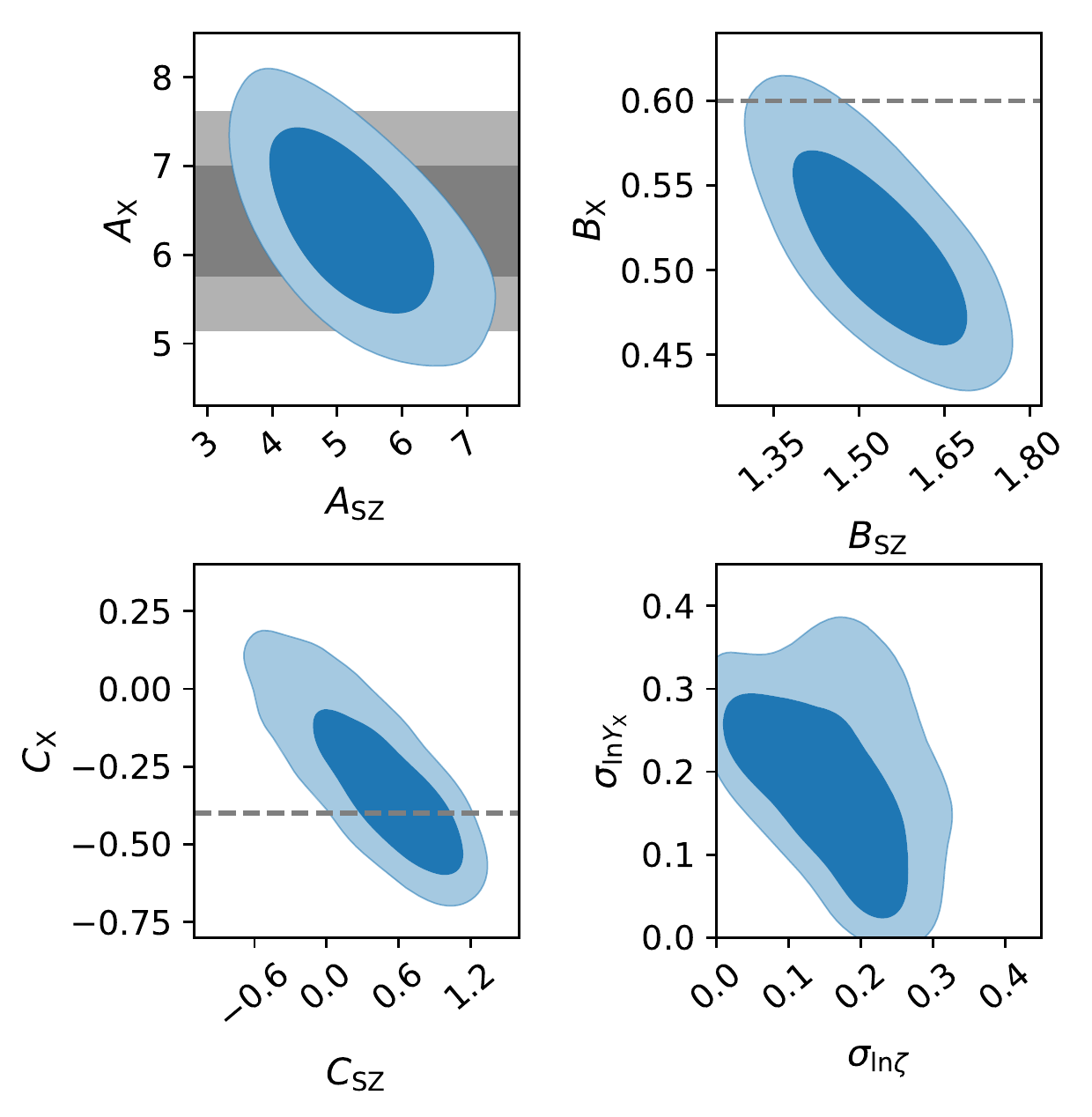}
  \caption{Our dataset is sensitive to the joint SZ-X-ray relation, which leads
  to correlations between the SZ and X-ray scaling relation amplitudes $A$ (top
  left), mass-slopes $B$ (top right), redshift evolutions $C$ (bottom left), and
  intrinsic scatters $\sigma$ (bottom right). We also show the external
  WL-informed prior on the X-ray amplitude \ayx\ applied in
  \citetalias{dehaan16}, and the self-similar expectations for the X-ray slope
  \byx\ and redshift evolution \cyx.}
  \label{fig:SZXraydegeneracy}
 \end{figure}

We compare our baseline results from SPTcl (SPT-SZ+WL+\yx) with the ones
obtained from the SPT-SZ+WL data combination, in which no X-ray follow-up data
are included. In this case, we apply an informative Gaussian prior to the SZ
scatter \sigmalnzeta. As in all of this work, no informative priors are applied
on the remaining three SZ scaling relation parameters and on the X-ray scaling
relation parameters. A figure showing constraints on all relevant parameters can
be found in Appendix~\ref{sec:appendixLCDM} (Fig.~\ref{fig:LCDM_WL_Yx_full},
compare blue and red contours) and Table~\ref{tab:constraints} summarizes
parameter constraints. Both data combinations, with and without X-ray data,
provide very similar constraints on cosmological and scaling relation
parameters. Without informative priors on the X-ray amplitude, mass-slope, or
redshift evolution the inclusion of X-ray data does not enable tighter
constraints. The use of X-ray data does, however, enables constraints on the SZ
and X-ray scatters \sigmalnzeta\ and \sigmalnyx, with flat priors applied to
both.

Note that our data set is sensitive to the SZ-to-X-ray relation. As defined in
Section~\ref{sec:massobsrel}, our model consists of two observable--mass
relations that each relate one observable to mass. This implies that the
amplitudes, mass-slopes, and redshift-evolutions of the two scaling relations
are degenerate, as shown in Fig.~\ref{fig:SZXraydegeneracy}. The degeneracy
between \sigmalnzeta\ and \sigmalnyx\ is particularly interesting: while the
marginalized posterior of either of both parameters has substantial mass near 0
scatter (see Fig.~\ref{fig:LCDM_WL_Yx_full}), the lower right panel of
Fig.~\ref{fig:SZXraydegeneracy} shows that 0 \emph{total scatter} is clearly
ruled out.

Our dataset is not able to constrain any of the coefficients describing the
correlated scatter among the observables. The visual impression of a constraint
in Fig.~\ref{fig:LCDM_WL_Yx_full} stems from the requirement that the matrix
describing the multi-observable scatter must be a valid non-degenerate
covariance matrix which prevents combinations of extreme correlation
coefficients.

\subsubsection{Constraints on X-ray Scaling Relation Parameters}
\label{sec:LCDMXray}

Without any informative priors on the X-ray scaling relation parameters, we can
use the SPTcl dataset to constrain the \yx--mass relation. The recovered
amplitude
\begin{equation}
\ayx=\LCDMclAyx
\end{equation}
is very close to the WL-informed prior \citep{hoekstra15, applegate14,
vonderlinden14a, mantz15} that was used in our previous cosmology analysis
\citepalias[$\ayx=6.38\pm0.61$,][]{dehaan16}. We constrain the redshift
evolution of the \yx--mass relation to
\begin{equation}
\cyx=\LCDMclCyx.
\end{equation}
The self-similar expectation $\cyx=-0.4$ is well within $1\sigma$. Our
measurement of the \yx\ scatter
\begin{equation}
\sigmalnyx=\LCDMclsigmalnyx
\end{equation}
is higher than, but consistent at the $1\sigma$ level with the prior
$0.12\pm0.08$ adopted in previous SPT analyses. It closely matches the measurement
$0.182\pm0.015$ from \cite{mantz16}, although with larger uncertainty.

The recovered \yx\ mass-slope
\begin{equation}
\byx=\LCDMclByx
\end{equation}
is lower than the self-similar evolution $\byx=0.6$ and the measurements
$\byx=0.57\pm0.03$ from \cite{vikhlinin09b} and
$\byx=1/(1.61\pm0.04)=0.621\pm0.015$ from \cite{mantz16}.\footnote{The scaling
relation in \cite{mantz16} is defined as a power law in mass, whereas we use a
power law in \yx.} From our data, the consistency of \byx\ with the self-similar
value is $p=\Byxpvalue$, corresponding to $\Byxsigma\sigma$. Our data constrain
\byx\ through its degeneracy with the SZ mass-slope \bsz\
(Fig.~\ref{fig:SZXraydegeneracy}), which in turn is constrained through the
process of fitting the cluster abundance against the HMF. This subject was
already discussed in \citetalias{dehaan16}, where a prior on \byx\ was adopted
from the measurement by \cite{vikhlinin09b}.

As a cross-check, and because other groups have used the X-ray gas mass as their
low-scatter mass proxy, we repeat the analysis replacing the \yx\ data with
\mgas\ measurements. We apply no informative priors on the four parameters of
the \mgas\ scaling relation of Eq.~\ref{eq:mgas}. We then analyze this
SPT-SZ+WL+\mgas\ dataset. The constraints on the SZ scaling relation parameters
and cosmology are very similar to the results from the fiducial SPT-SZ+WL+\yx\
analysis, and again we observe an X-ray mass-slope that disagrees with the
self-similar evolution. We measure
\begin{align}
\amg &=\LCDMAmg \label{eq:amg}\\
\bmg &=\LCDMBmg \\
\cmg &=\LCDMCmg \\
\sigmalnmgas &=\LCDMDmg.
\end{align}
This corresponds to a \LCDMBmgsigma\ preference for a slope that is steeper than
the self-similar expectation $\bmg=1$ or the measurement $\bmg=1.004\pm0.014$
from \cite{mantz16}. The measurement $\bmg=1.15\pm0.02$\footnote{
\cite{vikhlinin09b} use the functional form $\fgas = f_{\mathrm{gas},0} + \alpha
\ln M$. The mass dependence $\alpha$ is converted into a power-law exponent in
\cite{chiu18}. } from \cite{vikhlinin09b} is in between the two results, and is
$1\sigma$ low compared to ours. Because these slopes differ, we compare the
measurements of the gas fraction \amg\ at the pivot mass in our relation
$5\times10^{14}M_\odot/h_{70}$, where we obtain Eq.~\ref{eq:amg}. The mean gas
fraction at this mass is $0.128$ from \cite{mantz16} and $0.114$ from
\cite{vikhlinin09b}. Both values are contained within the $1\sigma$ range of our
measurement. Finally, as for \yx, our measurement of the redshift evolution
encompasses the self-similar evolution ($\cmg=0$) within $1\sigma$.

For an extensive discussion of the mass and redshift trends in the \mgas--mass
and \yx--mass relations for SPT selected clusters and how they compare to
previously published results, we refer the reader to two recent studies where SZ
based mass information was adopted using the posterior distributions of the SZ
$\zeta$--mass relation parameters presented in \citetalias{dehaan16}
\citep{chiu18,bulbul19}. \cite{bulbul19} used X-ray data from {\it XMM-Newton}
while we use data from \chandra; their recovered constraints on the X-ray mass
slopes and redshift evolutions are consistent with our findings at the $1\sigma$
level which confirms a consistent X-ray analysis. Here we note that most
measurements of X-ray scaling relations have been performed using samples at low
redshifts $z\lesssim 0.5$, and so it is of particular interest to examine the
mass slopes for the low redshift half of our sample.

\begin{figure}
  \includegraphics[width=\columnwidth]{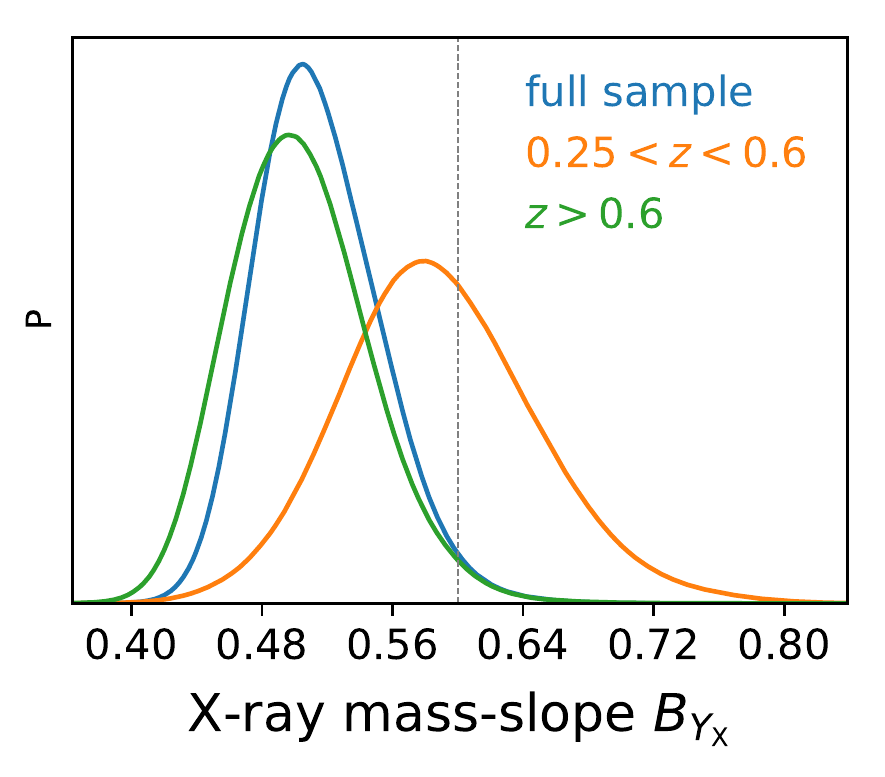}
  \caption{Constraint on the X-ray \yx\ slope \byx\ from the full sample, and
  from the low- and high-$z$ halves. The self-similar expectation $\byx=0.6$ is
  $\Byxsigma\sigma$ off the result from the full sample, but within $1\sigma$ of
  the low-$z$ result.}
  \label{fig:Bx}
\end{figure}

We therefore split our cluster sample (and all follow-up data) into two
subsamples above and below redshift of $z=0.6$, the median redshift of our
sample. Constraints on the most relevant parameters are shown in
Fig.~\ref{fig:LCDM_WL_Yx_full} in the Appendix, and Fig.~\ref{fig:Bx} shows the
constraints on \byx. Interestingly, the low-redshift subsample prefers a higher
value
\begin{equation}
\byx(0.25<z<0.6)=\LCDMByxloz
\end{equation}
that is closer to the self-similar evolution $\byx=0.6$. As expected, the value
obtained from the high-$z$ subsample
\begin{equation}
\byx(z>0.6)=\LCDMByxhiz
\end{equation}
is lower than the one obtained from the full sample. However, note that the
low-redshift and high-redshift constraints on \byx\ only differ with
$p=\LCDMByxlohipvalue$ ($\LCDMByxlohisigma\sigma$).

We perform the same splits in redshift using the SPT-SZ+WL+\mgas\ dataset. Here
as well, our measurement using the low-redshift subsample
\begin{equation}
\bmg(0.25<z<0.6)=\LCDMBmgloz
\end{equation}
is closer to the self-similar evolution, while the high-redshift half yields a
steeper slope
\begin{equation}
\bmg(z>0.6)=\LCDMBmghiz.
\end{equation}

To capture a possible redshift dependence of the slope of the X-ray scaling
relations, we analyzed models with an extended scaling relation model of the
form
\begin{equation}
\begin{split}
\ln \mathcal O_\textrm{X-ray} = & \ln A + B \ln \left(\frac{M_{500c} \, h_{70}}{5\times 10^{14}M_\odot}\right) \\
& + C \ln \left(\frac{E(z)}{E(0.6)}\right) \\
& + E \ln \left(\frac{E(z)}{E(0.6)}\right) \ln \left(\frac{M_{500c} \, h_{70}}{5\times 10^{14}M_\odot}\right)
\end{split}
\end{equation}
that allows for additional freedom and the mass- and redshift-dependences.
However, we do not observe any significant departure in $E$ from 0, in agreement
with \cite{bulbul19}.

\subsection{Constraints on the Sum of Neutrino Masses}

\begin{figure}
\includegraphics[width=\columnwidth]{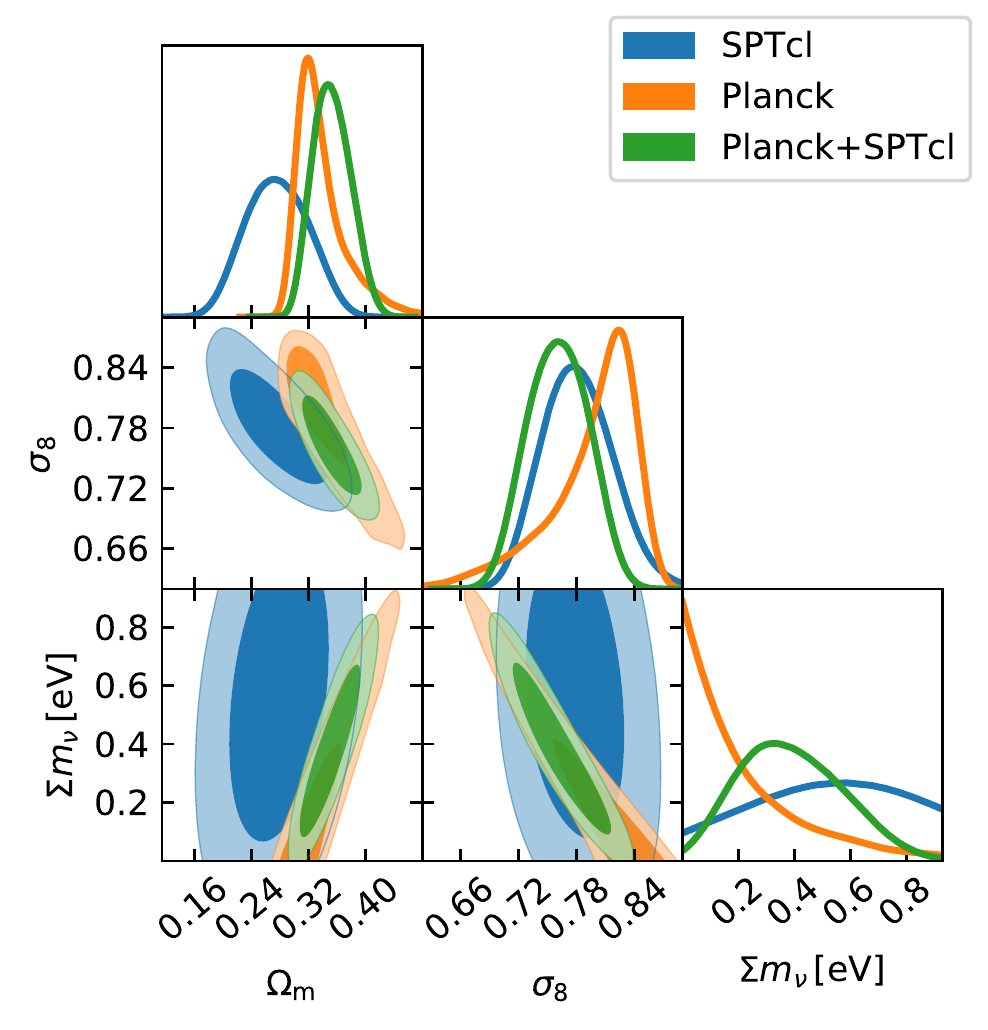}
\caption{\nuLCDM\ constraints on \Om, \sig, and \sumMnu. The SPTcl dataset
comprises (SPT-SZ+WL+\yx), $Planck$ uses TT+lowTEB. Note that the cluster data
constrain \Om\ and \sig\ almost independently of \sumMnu.}
\label{fig:LCDM_compare_Planck}
\end{figure}

Having quantified the consistency between our cluster dataset and \planck\ in
Section~\ref{sec:externalprobes}, we proceed and combine the two probes. The
SPTcl+\planck\ dataset yields
\begin{align}
\Om &= \LCDMclPlOm \\
\sig &= \LCDMclPlsig \\
\sigOmtwo &= \LCDMclPlsigOmtwo \\
\sumMnu &= \LCDMclPlmnu~\text{eV} \\
\sumMnu &< \LCDMclPlmnuNinetyfive ~\text{eV}\,\,(95\%\,\text{upper limit}).
\end{align}
Compared to constraints from \planck\ alone, the combination with SPTcl shrinks
the errors on \Om, \sig, and \sigOmtwo\ by $\LCDMclPlimproveOm\%$,
$\LCDMclPlimprovesig\%$, and $\LCDMclPlimprovesigOmtwo\%$. By breaking parameter
degeneracies (notably between \sig\ and \sumMnu, see
Fig.~\ref{fig:LCDM_compare_Planck}), the addition of cluster data to the primary
CMB measurements by \planck\ affects the inferred sum of neutrino masses. If
interpreted as a Gaussian probability distribution (i.e., ignoring the hard cut
$\sumMnu>0$), our joint measurement corresponds to a $\nuLCDMsummnusigma\sigma$
preference for a non-zero sum of neutrino masses.

\begin{figure}
  \includegraphics[width=\columnwidth]{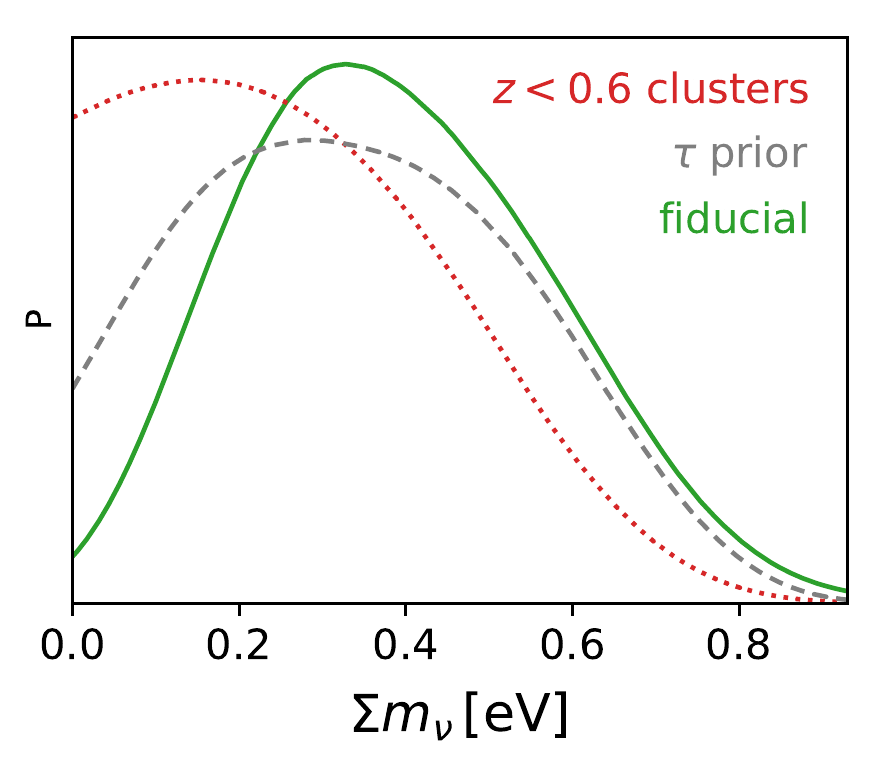}
  \caption{Constraints on \sumMnu\ from the joint analysis of SPTcl and \planck\
  data. Our fiducial analysis favors a non-zero sum of neutrino masses. However,
  when only using the low-redshift half $z<0.6$ of our cluster sample or when
  replacing \planck\ TT+lowTEB with \planck\ TT + a prior $\tau\sim\mathcal
  N(0.054,0.007^2)$ this preference diminishes.}
  \label{fig:summnu_diff}
\end{figure}

The \planck\ collaboration recently presented an updated analysis of primary CMB
anisotropies \citep{planck18-6}. Most notably, the optical depth decreased to
$\tau=0.054\pm0.007$. As the updated \planck\ likelihood code is not available
yet, we estimate the impact of the updated \planck\ analysis on our results and
especially our constraint on \sumMnu\ by analyzing the \planck\ 2015 TT data
(without lowTEB) with a prior on $\tau\sim\mathcal N(0.054,0.007^2)$. We analyze
the joint SPTcl+\planck\ TT+$\tau$prior dataset and obtain
\begin{equation}
\sumMnu = \LCDMclPlTTtaumnu~\text{eV}.
\end{equation}
The recovered constraint is lower than our fiducial constraint using the
(SPTcl+\planck\ TT+lowTEB) dataset and the $95\%$ credible interval runs against
the hard prior $\sumMnu=0$. The preference for a non-zero sum of neutrino masses
reduces to $\nuLCDMtausummnusigma\sigma$. We caution the reader that this result
is only preliminary due to the way it depends on the prior on $\tau$ that we
adopted. The full analysis will require analyzing our cluster sample jointly
with the latest \planck\ analysis.

We explain the shift in \sumMnu\ toward lower values as follows. In \LCDM, the
relationship between $A_s$ and \sig\ is essentially fixed. However, in \nuLCDM,
the additional degree of freedom \sumMnu\ allows for different values of \sig\
at a fixed $A_s$. In any joint analysis of \planck+low-redshift
growth-of-structure-probe as SPTcl, \sumMnu\ is constrained to accommodate the
\planck\ measurement of $A_s$ with the low-redshift measurement of \sig. As has
been pointed out many times, the \planck15 measurement of $A_s$ implies a higher
\sig\ in \LCDM\ than obtained from local measurements, which leads to an
apparent  detection of \sumMnu\ in \nuLCDM. Meanwhile, CMB temperature
fluctuations are sensitive to the combination $A_s e^{-2\tau}$---i.e., $A_s$ and
$\tau$ are positively correlated in TT parameter constraints---so imposing a
$\tau$ prior with a lower central value results in a lower inferred value of
$A_s$. In \LCDM, this shifts the \planck-inferred \sig\ to lower values.
Finally, when analyzing \planck\ TT+$\tau$+SPTcl in \nuLCDM, \sig\ is dominated
by the local constraint from SPTcl, and the lower $A_s$ implies that \sumMnu\
need not be as high as in our fiducial analysis.

We further test the impact of using only the low-redshift half of our cluster
sample. The SPTcl($0.25<z<0.6$)+\planck\ dataset yields
\begin{equation}
\sumMnu = \LCDMclzsmpsixPlmnu~\text{eV}.
\end{equation}
The probability distribution in \sumMnu\ runs against the hard prior $\sumMnu>0$
which shifts the mean recovered value away from the mode; the 68\% credible
interval starts at $\sumMnu=0$. In conclusion, all preference for a non-zero sum
of neutrino masses vanishes when only considering the low-redshift half of our
cluster sample. Fig.~\ref{fig:summnu_diff} shows the constraints on \sumMnu\ as
obtained in our fiducial analysis, the analysis with the $\tau$ prior and the
analysis where we only use the low-redshift cluster data.

\begin{figure}
  \includegraphics[width=\columnwidth]{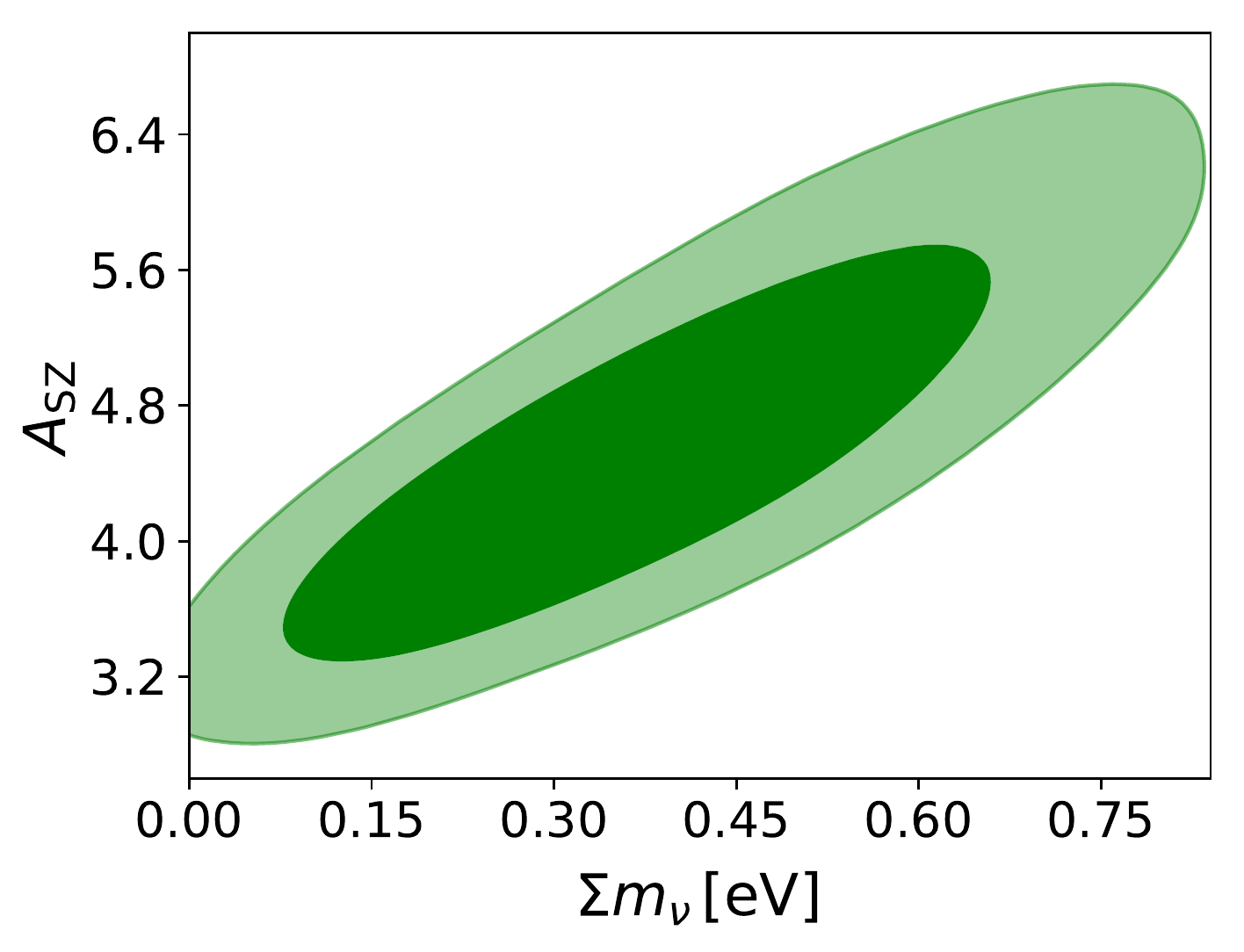}
  \caption{Parameter correlation between the sum of neutrino masses \sumMnu\ and
  the amplitude of the SZ observable--mass relation \asz\ for the SPTcl+\planck\
  dataset. An improved cluster mass calibration will enable tighter constraints
  on neutrino properties.}
  \label{fig:summnu_Asz}
\end{figure}

The sum of neutrino masses is degenerate with the amplitude of the SZ scaling
relation \asz\ with a correlation coefficient
$\rho_{\asz-\sumMnu}=\LCDMcorrAszMnu$, see Fig.~\ref{fig:summnu_Asz}. Therefore,
an improved (WL) mass calibration will improve the constraints on \sumMnu. Also
note that the effect of massive neutrinos on the HMF depends (weakly) on mass
and redshift \citep{ichiki11}. Therefore, an improved mass calibration covering
the entire cluster sample will in principle allow for measurements of the sum of
neutrino masses from clusters alone.

\subsection{\nuwCDM\ Cosmology}
\label{sec:wCDM}

\begin{figure*}
  \includegraphics[width=\textwidth]{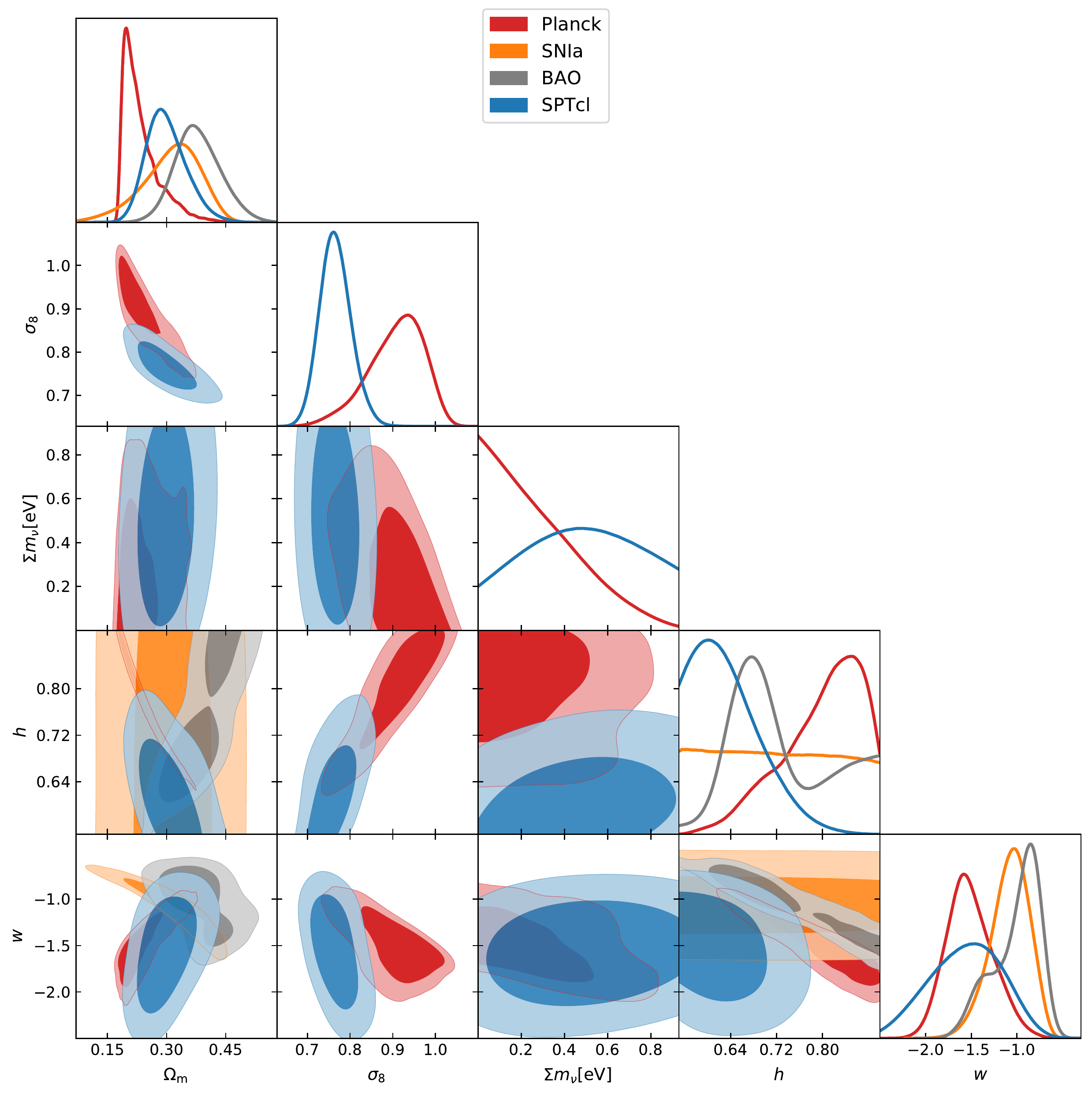}
  \caption{Constraints on \nuwCDM\ from SPT clusters, \planck, BAO, and SNIa.
  The $95\%$ credibility contours all overlap. The biggest differences appear
  between SPTcl and \planck\ in the \sig\ and $h$ parameters.}
  \label{fig:wCDM_single}
\end{figure*}

\begin{figure*}
  \includegraphics[width=\textwidth]{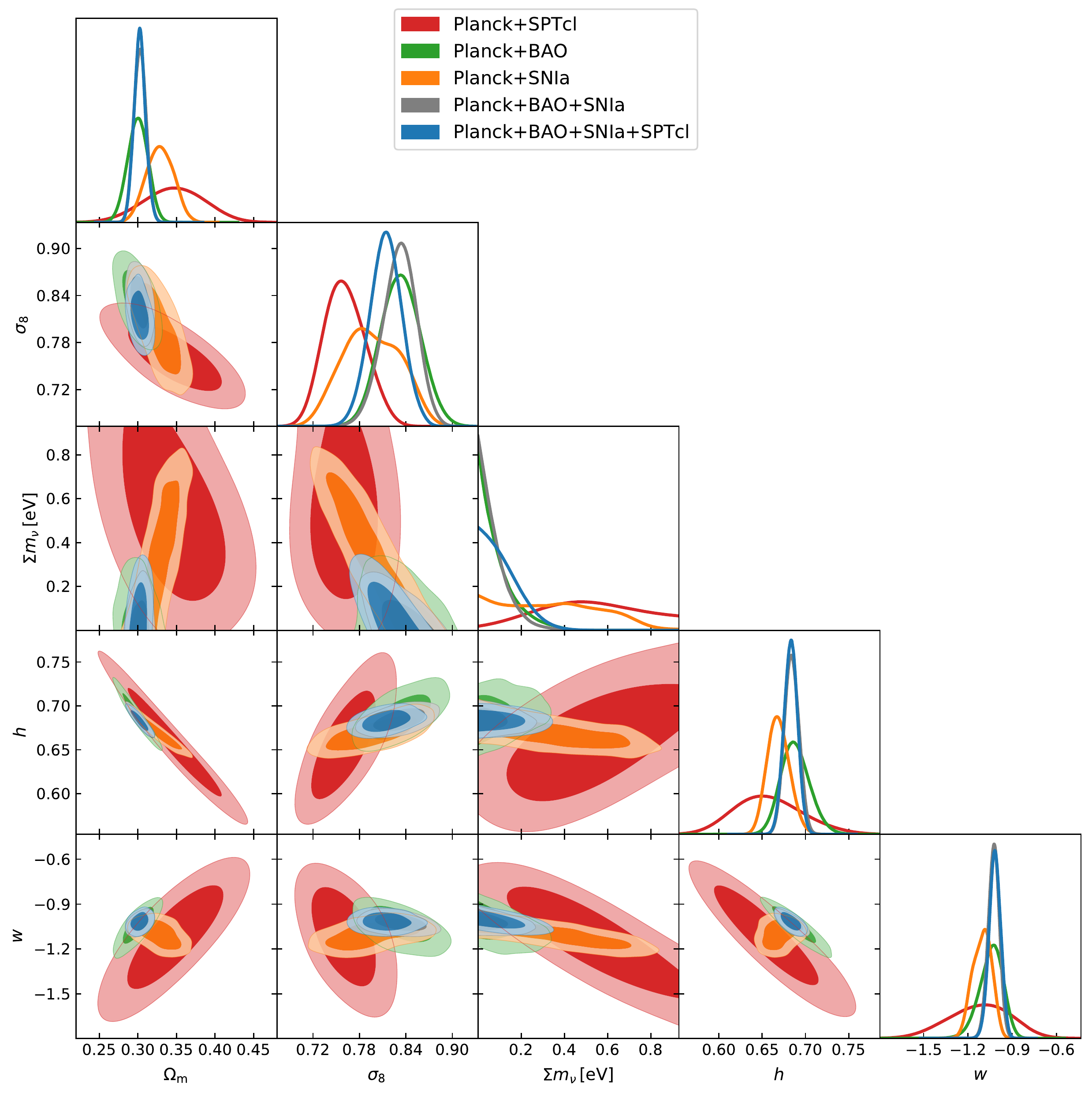}
  \caption{Constraints on \nuwCDM\ from joint analyses of \planck\ with SPTcl,
  BAO, or SNIa. We also show \planck+BAO+SNIa and the full joint analysis
  \planck+BAO+SNIa+SPTcl. When combining with \planck, our cluster dataset does
  not contribute as much additional information as do the other two external
  probes BAO and SNIa.}
  \label{fig:wCDM_combined}
\end{figure*}

We consider an extension to modeling dark energy as a cosmological constant by
allowing for an equation of state parameter $w$ that is different from $w=-1$.
This modification impacts the expansion history of the Universe $E(z)$ and the
growth of structure; both affect the cluster abundance. Therefore, as noted in,
e.g., \cite{haiman01}, measuring the abundance over a range of redshifts allows
for a measurement of $w$. Using our SPTcl dataset we obtain
\begin{align}
w &= \wCDMclw \label{eq:w}\\
\Om &= \wCDMclOm\\
\sig &= \wCDMclsig,
\end{align}
as shown by the blue contours in Fig.~\ref{fig:wCDM_single}. Constraints on
scaling relation parameters can be found in Table~\ref{tab:constraints}. The
consistency of our recovered constraint on $w$ with a cosmological constant
$w=-1$ has a $p$-value \wCDMpvalue\ ($\wCDMsigma\sigma)$. Note that the SPTcl
contours in the $\Om-\sig-w$ space close.

Our constraint on $w$ is in general agreement with the result obtained from the
SPT-SZ+\yx+X-ray priors data combination $w=-1.28\pm0.31$ as presented in
\citetalias{dehaan16}. In that earlier analysis, informative (Gaussian) priors
were applied on the scaling relation parameters \asz, \bsz, \csz, \sigmalnzeta,
\ayx, \byx, \cyx, \sigmalnyx, whereas we marginalize over flat priors and use
our internal WL mass calibration. However, even when analyzing the same data
combination used in \citetalias{dehaan16} (without WL data) and applying the
same priors, our analysis pipeline gives a more negative value of $w=\wCDMwYx$.
As described in Section~\ref{sec:validation}, we have extensively tested our
analysis pipeline, including tests against mock catalogs with input values of
$w\neq-1$. The analysis pipeline used in \citetalias{dehaan16} was not subjected
to that test. Using our internal WL mass calibration shifts the constraints on
$w$ toward even more negative values. Finally, the cluster photometric redshifts
were updated since the \citetalias{dehaan16} analysis (see
Section~\ref{sec:SPTsample}), with the net impact being a shift in $w$ toward
less negative values of similar magnitude to the shift due to our WL mass
calibration. In the end, some of these shifts in $w$ partially cancel out, and
the final constraint we present here is $0.7\sigma$ low in comparison to that in
\citetalias{dehaan16}.

We proceed and analyze the joint SPTcl+\planck\ dataset. The cluster data break
some of the \planck\ parameter degeneracies shown in
Fig.~\ref{fig:wCDM_single} and we measure
\begin{align}
w &= \wCDMclPlw \label{eq:w_clPl}\\
\Om &= \wCDMclPlOm\\
\sig &= \wCDMclPlsig\\
\sumMnu &= \wCDMclPlmnu~\mathrm{eV}.
\end{align}
Interestingly, while the individual constraints on $w$ are both centered on
$w\approx-1.5$, the joint analysis provides a constraint that is offset closer
toward $w=-1$. This is due to the different orientations of the $w-\sig$
degeneracies in Fig.~\ref{fig:wCDM_single} which overlap close to $w=-1$.
Compared to the results obtained in \nuLCDM, the constraints on \sig\ and
\sigOmtwo\ do not degrade. However, the constraining power on the remaining
cosmological parameters weakens (see Table~\ref{tab:constraints}).

Fig.~\ref{fig:wCDM_single} further shows the constraints obtained from BAO and
SNIa. Neither of the two are affected by \sig\ and \sumMnu.\footnote{We note an
unexpected shape of the BAO posterior on $h$, with a peak at $h\approx0.68$ and
a rising tail toward the edge of the prior $h<0.9$. This is caused by the
subsamples of BAO providing different results: The 6dF+SDSS posterior peaks at
$h\approx0.68$ and exhibits an approximately flat, non-zero tail in the range
$0.8<h<0.9$. The posterior from BOSS increases monotonically throughout the
entire allowed range in $h$ and peaks at $h=0.9$. Therefore, the joint BAO
dataset peaks at the 6dF+SDSS location but then rises again toward $h=0.9$ due
to the BOSS constraint.} However, they both exhibit narrow parameter
degeneracies that cut through the region of parameter space that is allowed by
\planck. Therefore, the joint analyses of \planck+BAO and \planck+SNIa allow for
constraints on \nuwCDM\ that are tighter than the ones from \planck+SPTcl (see
Fig.~\ref{fig:wCDM_combined}).

Finally, we analyze the joint \planck+BAO+SNIa+SPTcl dataset (see constraints in
Table~\ref{tab:constraints}). In comparison to \planck+BAO+SNIa, the addition of
the SPTcl dataset leads to a shift $\Delta\sig=\nuwCDMshiftsig$. The
constraints on \Om, $h$, and $w$ are negligibly affected. However, note that the
$95\%$ upper limit on \sumMnu\ from \planck+BAO+SNIa increases by
$\mnuupperlimitshift\%$ when adding SPTcl. A similar effect was seen in the DES
3x2 pt analysis \citep{des18-main}, where the upper limit on \sumMnu\ from
\planck+BAO+SNIa increased by a similar amount when adding the DES data. Both
effects are due to the lower clustering amplitude measured by SPTcl and DES
relative to the prediction by \planck+BAO+SNIa.

\subsubsection{\nuwCDM: Robustness of our Results to Data Cuts}

In the Appendix (Fig.~\ref{fig:wCDM_full}), we show the parameter constraints
that we recover when cutting our cluster sample in half at redshift $0.6$, or when
choosing a higher SZ selection threshold $\xi>6.5$. There are no significant
departures from our fiducial results for any data subset. However, both the
low-redshift half of the data and the subsample above $\xi>6.5$ yield
constraints on $w$ that are closer to the cosmological constant $w=-1$:
\begin{align}
w(0.25<z<0.6) &= \wCDMwloz \\
w(\xi>6.5) &= \wCDMwxigtr.
\end{align}
Conversely, the high-redshift half of the data gives
\begin{equation}
w(z>0.6) = \wCDMwhiz.
\end{equation}
We note that the constraints on $w$ from the full sample is quite similar to
this constraint from the high-redshift half of the data.

Fig.~\ref{fig:wCDM_full} further shows a strong degeneracy between $w$ and the
redshift evolution parameters of the scaling relations \csz\ and \cyx. To
tighten the dark energy constraints in future analyses it will therefore be
important to improve the mass calibration over the entire redshift range of the
cluster sample.

\subsection{Growth of Structure: Measuring $\sigma_8(z)$}
\label{sec:sigma8ofz}

\begin{figure*}
\includegraphics[width=\textwidth]{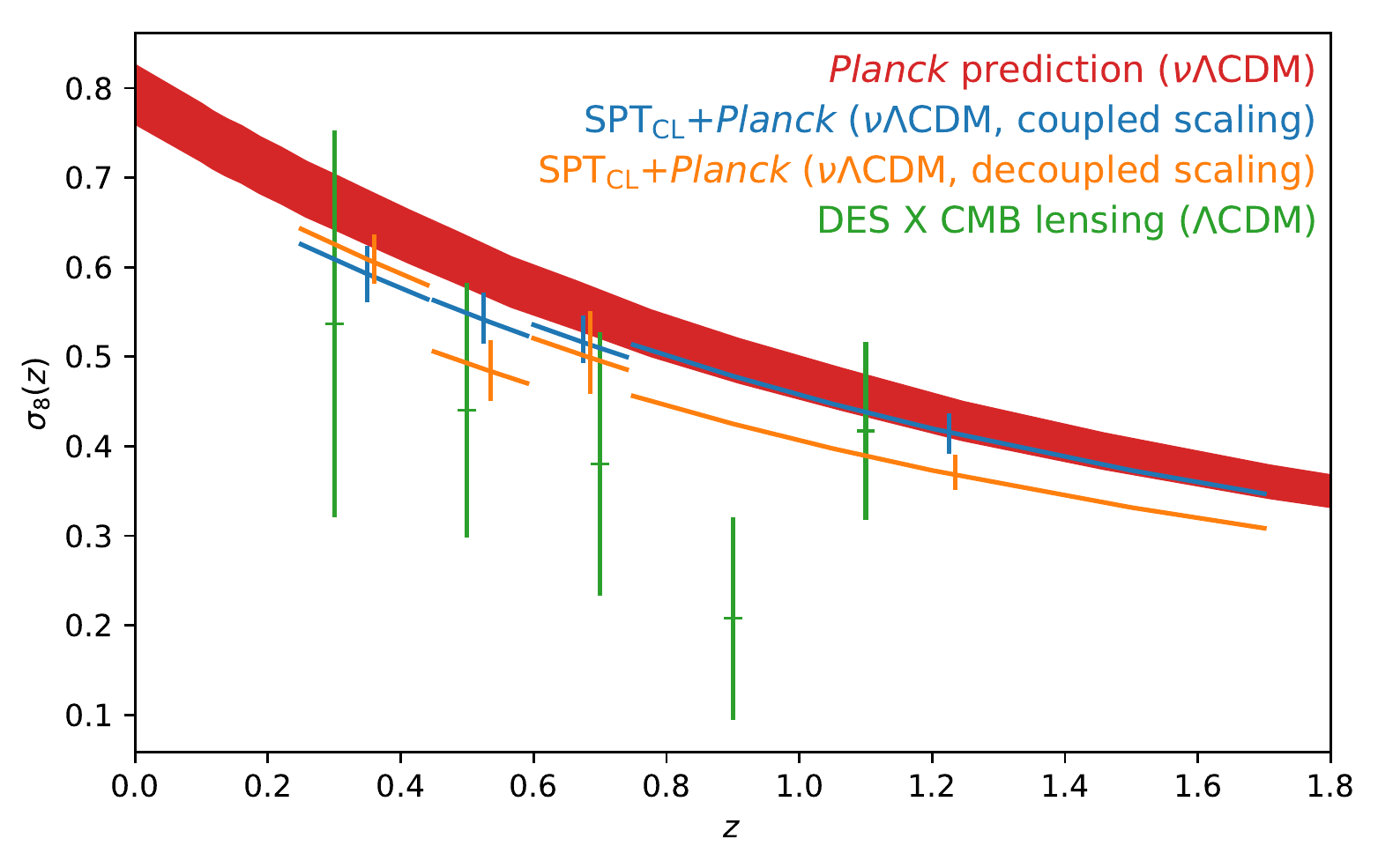}
\caption{The evolution of \sig\ as a function of redshift. The \emph{red band}
shows the $1\sigma$ interval of the prediction obtained from \planck\ in the
\nuLCDM\ cosmology. The \emph{blue data points} are obtained in a joint
SPTcl+\planck\ analysis, where \sigz\ is constrained only by the cluster sample.
\emph{Orange data points} are obtained from a similar analysis that allows for
more freedom in the SZ scaling relation (see Section~\ref{sec:sigma8ofz}). The
nearly horizontal error bars on the blue and orange data points indicate the
extent of the redshift bins and are shaped to follow the evolution of \sig\ in
the \nuLCDM\ model. For comparison, \emph{green data points} show constraints
from the cross-correlation of the galaxy density in the Dark Energy Survey
Science Verification data with CMB lensing from SPT \citep{giannantonio16}.}
\label{fig:sigma8_z}
\end{figure*}

We consider another extension to \LCDM\ where we do not alter the background
expansion, but change the growth of structure. Clusters have been used to
constrain modified structure growth by, for example, fitting for the growth
index $\gamma$, which is defined by the relation $d \ln \delta/d\ln a \equiv
\Om^\gamma(a)$ \citep[e.g.,][]{peebles80}. A value of $\gamma\approx0.55$
corresponds approximately to the growth rate in \LCDM, and clusters allow for
constraints on $\gamma$ at the $\sim40\%$ level \citep[e.g.,][]{rapetti13,
bocquet15, mantz15}.

Instead of modeling linear deviations from GR via the growth index, we pursue a
different route and constrain the growth of structure by directly measuring the
linear amplitude of the density fluctuations, $\sig$, as a function of redshift.
We can then compare the measured \sigz\ with predictions from \nuLCDM, \nuwCDM,
and more exotic models. This approach is non-parametric in that it does not
assume a specific description for modified growth of structure, but rather
assumes a \nuLCDM\ model (with its parameters allowed to vary) within each
redshift bin.

We start from the \nuLCDM\ model and modify the amplitude of the linear matter
power spectrum $P(k,z)$ within different redshift bins. We introduce an
additional model parameter $\sigma_8(z_i)$ in each bin and normalize $P(k,z)$
within each redshift bin $i$ to match $\sigma_8(z_i)$. The HMF is then computed
from the modified $P(k,z)$ in the usual way. We define four redshift bins such
that all bins contain approximately equal numbers of SPT clusters. We choose bin
limits ($z=0.25, 0.45, 0.6, 0.75, 1.7$). We include \planck\ primary CMB data in
the fit. By construction, in our model the \planck\ data only constrain the
background cosmology (expansion history $E(z)$), but do not contribute to
constraining $\sig(z)$. For simplicity, we do not use any X-ray data here and
thus use the joint SPT-SZ+WL+\planck\ dataset; we apply the simulation prior on
the SZ scatter.

We explore two scenarios: i) We assume our fiducial SZ scaling relation model
across the entire redshift range. This means that the mass calibration will be
correlated across the four redshift bins. ii) We additionally introduce
independent normalizations of the SZ scaling relation $A_{\mathrm{SZ},i}$ in
each redshift bin $i$. This way, the amplitude of the SZ scaling relation is
independently determined in each redshift bin (up to the shared WL systematics
that are, however, sub-dominant here given the low number of clusters with WL
constraints). We call the first scenario ``coupled'', and the second
``decoupled'', in reference to the treatment of the normalization of the SZ
observable--mass relation.

\begin{deluxetable*}{lccc}
\tablecaption{\label{tab:sigma8ofz}
  Constraints on \sig\ and the SZ scaling relation normalization \asz\ as a
  function of redshift, measured in four redshift bins. In the \emph{coupled
  analysis} we assume a single \asz\ parameter. In the \emph{decoupled analysis}
  we fit for \asz\ separately in each redshift bin; this decorrelates the
  measurements of $\sig(z)$ as evidenced by $\rho(\sig(z))$.}
\tablehead{\colhead{Parameter} & \colhead{Coupled analysis} & \colhead{Decoupled
analysis} & \colhead{\planck\ (TT+lowTEB)}}
\startdata
\asz & $5.40^{+0.80}_{-1.15}$ & \nodata & \nodata \\
$\asz(0.25<z<0.45)$ & \nodata & $4.90^{+0.88}_{-1.09}$ & \nodata \\
$\asz(0.45<z<0.6)$ & \nodata & $10.29^{+2.56}_{-4.23}$ & \nodata \\
$\asz(0.6<z<0.75)$ & \nodata & $7.29^{+1.66}_{-4.47}$ & \nodata \\
$\asz(0.75<z<1.7)$ & \nodata & $10.63^{+2.19}_{-3.08}$ & \nodata \\[6pt]
$\sig(z=0.35)$ & $0.592\pm0.031$ & $0.609\pm0.028$ & $0.656\pm0.029$ \\
$\sig(z=0.525)$ & $0.543\pm0.029$ & $0.484\pm0.034$ & $0.597\pm0.028$ \\
$\sig(z=0.675)$ & $0.519\pm0.026$ & $0.505\pm0.046$ & $0.555\pm0.026$ \\
$\sig(z=1.225)$ & $0.415\pm0.023$ & $0.371\pm0.020$ & $0.432\pm0.021$ \\[6pt]
$\rho(\sig(z))$ & $0.55-0.60$ & $0.06-0.12$ & \nodata
\enddata
\end{deluxetable*}

Constraints on \asz\ and $\sig(z)$ for the coupled and decoupled analyses are
shown in Table~\ref{tab:sigma8ofz}. In the coupled analysis, the four
measurements of $\sig(z)$ are quite correlated with correlation coefficients
$\rho(\sig(z))=\coupledrho$ because they are limited by the uncertainty in the
observable--mass scaling relation that is shared across the entire redshift
range. In the decoupled analysis, however, the $\sig(z)$ parameters are much
less correlated ($\rho(\sig(z))=\decoupledrho$) as mass calibration in each bin
is done almost independently, and each $\sig(z_i)$ is mostly degenerate with the
corresponding normalization parameter $\asz(z_i)$. As expected, the decoupled
analysis leads to weaker constraints.

In Fig.~\ref{fig:sigma8_z}, we show measurements of \sig\ as a function of
redshift. The red bands shows the \emph{prediction} for $\sig(z)$ assuming
\nuLCDM\ and \planck\ cosmological parameters. Blue and orange data points show
\emph{measurements} of $\sig(z)$ using our clusters (with \planck\ priors on the
background cosmology). The cluster measurements are all slightly lower than the
predictions using \planck\ data, which simply reflects the difference in \sig\
discussed above for the \nuLCDM\ model (see Fig.~\ref{fig:LCDM_compare_Planck}).
We emphasize that this offset is roughly constant throughout the entire range in
redshift. In particular, the two bins above $z>0.6$ that are leading to some
shifts in cosmology and scaling relations as described in earlier sections do
not seem to provide constraints that are qualitatively different from those
obtained from the low-redshift bins.

Our measurements of $\sig(z)$ are limited by the determination of \asz,
especially in the ``decoupled'' analysis. The three low-redshift bins will
benefit from including cluster WL data from the Dark Energy Survey (Paulus et
al., in prep.). The highest-redshift bin can only be improved with deep,
high-resolution WL data, e.g., from the \hst, or with lensing information from
the CMB \citep[e.g.,][]{baxter18}. On the other hand, our cluster sample
together with this technique allow us to place competitive constraints on the
growth of structure over a wide range in redshifts.

\subsection{Implications for SZ-Based Cluster Halo Masses}
\label{sec:masses}

For the \nuLCDM\ and \nuwCDM\ analyses discussed above,
Table~\ref{tab:constraints} also presents constraints on the SZ scaling relation
parameters. These, together with Eqs.~\ref{eq:zetaM} and \ref{eq:xi2zeta} allow
one to compute mass estimates $P(M_{500c}|\,\xi,z,\vec p)$ for each cluster in
our sample. Moreover, the scaling relation parameter constraints provide another
point of comparison with past analyses.

The results in Table~\ref{tab:constraints} exhibit a range of parameters across
the six different analyses, but in no case are the parameter differences
statistically significant. This indicates that the best estimates of the cluster
masses are consistent among the different combinations of data and within the
different cosmological models. As an example, the addition of the \planck\
dataset as an external prior leads to preferred values of the amplitude
parameter \asz\ that are lower, corresponding to $\sim\massshiftL\%$ and
$\sim\massshiftw\%$ higher masses at the pivot in \nuLCDM\ and \nuwCDM,
respectively. These mass shifts are smaller than those presented by
\citet{bocquet15}, where the impact of external priors was first discussed.
Interestingly, the redshift slope \csz\ prefers higher values in the \nuwCDM\
model, which corresponds to high-redshift masses that are smaller relative to
clusters with the same $\xi$ at low redshift. In the \nuwCDM\ model these same
values of $\csz\sim1$ are preferred with or without \planck\ priors, but shift
back to a lower value when BAO+SNIa constraints are added. In comparison, in the
results for \LCDM\ presented in Table 3 of \citetalias{dehaan16}, the amplitude
parameter for the SPTcl+\planck+BAO analysis was $\asz=3.53\pm0.27$, which is
significantly lower than the values presented here. Note, however, that
massive neutrinos were not marginalized over in the baseline analysis in
\citetalias{dehaan16}.

Given the consistency in the implied masses across all six analyses presented
here, we adopt the \nuLCDM\ results for the baseline SPTcl dataset in
calculating mean masses and mass uncertainties (Table~\ref{tab:constraints}
column 3). The mass uncertainties include the $\xi$ measurement and intrinsic
scatter uncertainties (together these correspond to $\sim20\%$ uncertainty for a
cluster near our selection threshold) as well as marginalization over the
posterior parameter distributions for \asz, \bsz, \csz\ and \sigmalnzeta\ and
over the cosmological parameters (this corresponds to an additional $\sim15$\%
uncertainty due to the remaining uncertainties in the mass calibration of the
SPT-SZ sample). These masses are calculated by sampling the distribution
\begin{equation}
P(M|\xi, z, \vec p) = \iint dM d\zeta P(\xi|\zeta) P(\zeta|M, z, \vec p) P(M| z, \vec p)
\end{equation}
at each step in the likelihood analysis.

Table~\ref{tab:catalog} contains a list of all cluster candidates in our sample
with the associated sky location, SPT detection significance $\xi$, redshift and
halo mass $M_\mathrm{500c}$. In addition, we present the mass $M_\mathrm{200c}$
for each system, assuming the concentration--mass relation from \citet{duffy08}.

\section{Summary and Outlook}

We present an analysis of the SPT-SZ cluster sample, supplemented with optical
WL data and X-ray \yx\ measurements. We set up a self-consistent analysis
framework in which cosmology, scaling relations, a possible correlated scatter
among cluster mass proxies, and other nuisance parameters are fit
simultaneously. Within this framework, the WL data is used to constrain the
normalization of the observable--mass relations (at various redshifts and
cluster masses). We use numerical simulations to calibrate the relation between
the unobserved, true halo mass, and the observed radial WL shear profiles. Wide,
non-informative priors are assumed on the parameters of the SZ and X-ray scaling
relations. At present, our mass calibration is limited by the number of clusters
with WL data; the systematic uncertainties in the WL analysis are sub-dominant
with $5.6\%$ in mass for our ground-based data and $9.2-9.4\%$ for the HST sample
\citepalias{schrabback18b,dietrich19}.

Our main findings are:
\begin{itemize}
\item Assuming simulation-based priors on the relation between true mass and WL
mass \mwl, we are able to simultaneously fit for cosmology (constraining \Om,
\sig, $w$) while constraining the amplitudes, mass-slopes, redshift-evolutions,
and intrinsic scatter of the SZ and X-ray observable--mass relations. We
marginalize over flat priors on \Omnuhh, \Ombhh, $n_s$, $h$ which are not
constrained from cluster data alone.

\item Assuming the \nuLCDM\ model, our cluster-based constraint on
$\sigOmtwo=\LCDMclsigOmtwo$ is lower than the one obtained from primary CMB
fluctuations by \planck. The agreement between the two measurements is
$p=\nuLCDMclPlsigOmtwopvalue$, or $\nuLCDMclPlsigOmtwosigma\sigma$.

\item We constrain the redshift evolution of the X-ray \yx--mass relation to
$\cyx=\LCDMclCyx$ and the redshift evolution of the \mgas--mass relation to
$\cmg=\LCDMCmg$. The self-similar evolution $-0.4$ for \yx--mass and $0$ for
\mgas--mass is encompassed in the $1\sigma$ interval in both cases.

\item We find the mass-dependence of the X-ray \yx--mass relation
$\byx=\LCDMclByx$ to be steeper than the self-similar expectation $\byx=0.6$
with a $p$-value of $p=\Byxpvalue$ ($\Byxsigma\sigma$). Interestingly, this
difference is resolved when we only consider the low-redshift half of our sample
at $z<0.6$, where we measure $\byx=\LCDMByxloz$. Conversely, the high-redshift
half of our sample favors a steeper slope ($\byx=\LCDMByxhiz$), see also
Fig.~\ref{fig:Bx}. The slope of the \mgas--mass relation $\bmg=\LCDMBmg$ is also
steeper than the self-similar evolution $\bmg=1$. Here as well, the measurement
of $\bmg$ at low redshift below $z=0.6$ is closer to the self-similar value
($\bmg=\LCDMBmgloz$) than the high-redshift measurement ($\bmg=\LCDMBmghiz$).

\item The joint dataset combining our clusters and primary CMB measurements from
\planck\ allows for a constraint on the sum of neutrinos masses
$\sumMnu=\LCDMclPlmnu$~eV ($\sumMnu<\LCDMclPlmnuNinetyfive$~eV ($95\%$ C.L.)).
This preference for a non-zero sum of neutrino masses diminishes when combining
\planck\ with only the low-redshift ($z<0.6$) half of our cluster sample or when
adopting a lower value of $\tau$ as suggested by \cite{planck18-6}. Due to
parameter degeneracies, an improved cluster mass calibration will directly
translate into tighter constraints on neutrino masses.

\item Our constraint on $w=\wCDMclw$ is somewhat lower than a cosmological
constant, with $p=\wCDMpvalue$ ($\wCDMsigma\sigma$). The SPTcl contours in the
$\Om-\sig-w$ space are closed, see Fig.~\ref{fig:wCDM_single}. This reflects the
fact that our cluster sample is able to constrain the three parameters
simultaneously. When only considering the high-redshift $z>0.6$ subsample, we
obtain $w=\wCDMwhiz$, whereas we obtain a less negative constraint $w=\wCDMwloz$
from the low-redshift subsample at $0.25<z<0.6$.

\item We employ a new approach to measuring the linear growth of structure using
clusters. This allows us to track the evolution of structure growth since
redshift $z\sim1.7$. Fig.~\ref{fig:sigma8_z} shows that structure formation
evolved in agreement with the \nuLCDM\ prediction, although with a somewhat
lower amplitude than predicted assuming cosmological parameters from \planck.

\end{itemize}

The validity of our cluster-based constraints relies on an accurate prediction
of the HMF throughout the entire parameter space considered. However, the HMF
fit by \cite{tinker08}, is calibrated using $N$-body simulations for cosmologies
that are close to WMAP results, and the extrapolation to other cosmologies is
performed assuming the universality of the HMF. Ongoing analyses of cosmological
simulations will provide accurate predictions of the HMF for a much broader
range of cosmologies \citep[][Bocquet et al., in prep.]{heitmann16,
mcclintock19b}.

We discuss the departure from self-similarity of the X-ray \yx\ and \mgas\
mass-slopes. There is a suggestion of an evolution of the \yx\ mass slope with
redshift, where it exhibits more self-similar results in the low-redshift half
cluster sample. Similar results have been presented in the previous SPT
cosmology analysis \citepalias{dehaan16} as well as in X-ray observable--mass
scaling relation studies that rely on SZ based cluster masses
\citep{chiu16,chiu18,bulbul19}, where masses are calculated using the mass
calibration results from previous SPT cluster cosmology analyses
\citep[][dH16]{bocquet15}. This could be a sign that X-ray scaling relations
depart from self-similarity in this mass and redshift range \citep[e.g., the ICM
mass fraction varies with cluster mass as shown first in][]{mohr99}, or there
could be additional effects not captured by our model that affect e.g., the SZ
scaling relation or selection. Larger SZ-selected cluster samples and more
extensive follow-up data are necessary to discern these effects.

In upcoming analyses, we will expand our SPT-SZ cluster sample with data from
SPTpol. This will both increase our sample of high-mass clusters, and push down
to lower cluster masses in the deeper fields of the survey. At the same time, it
is important to pursue our WL campaign at all redshifts covered by our sample.
Indeed, the strategic overlap with the DES \citep{DES16} will allow for a robust
mass-calibration at moderate redshifts \citep{melchior17, stern19,
mcclintock19a}. To exploit the full potential of the SPT cluster sample, it will
be crucial to also tighten the WL mass constraints at higher redshifts. At
intermediate redshifts this can be achieved with deep ground-based
$K_\mathrm{s}$ imaging \citep{schrabback18a}, but at high redshifts $z>1$ these
measurements critically require additional HST observations or ultimately the
datasets from Euclid \citep{laureijs11} and LSST \citep{ivezic08}. With the
current and next generation of high-resolution CMB experiments such as SPT-3G
\citep{benson14, bender18}, Advanced ACTpol \citep{debernardis16}, or CMB-S4
\citep{abazajian16}, CMB lensing will provide another means of accurate mass
calibration out to redshifts well beyond $z\sim1$.

\acknowledgments

We thank Holger Israel for his feedback on the manuscript and Joe Zuntz for
support with \textsc{CosmoSIS}.
This work is partially based on observations made with the NASA/ESA \hst, using
imaging data from the SPT follow-up GO programs 12246 (PI: C.~Stubbs) and 12477
(PI: F.~W.~High), as well as archival data from GO programs 9425, 9500, 9583,
10134, 12064, 12440, and 12757, obtained via the data archive at the Space
Telescope Science Institute, which is operated by the Association of
Universities for Research in Astronomy, Inc. under NASA contract NAS 5-26555.
The Munich group acknowledges the support by the DFG Cluster of Excellence
``Origin and Structure of the Universe'', the Transregio program TR33 ``The Dark
Universe'', the Max-Planck-Gesellschaft Faculty Fellowship Program, and the
Ludwig-Maximilians-Universit{\"a}t Munich.
Work at Argonne National Laboratory was supported under the U.S. Department of
Energy contract DE-AC02-06CH11357.
TS acknowledges support from the German Federal Ministry of Economics and
Technology (BMWi) provided through DLR under projects 50 OR 1210, 50 OR 1308, 50
OR 1407, 50 OR 1610, and 50 OR 1803.
The Stanford/SLAC group acknowledges support from the U.S. Department of Energy
under contract number DE-AC02-76SF00515, and from the National Aeronautics and
Space Administration (NASA) under Grant No. NNX15AE12G, issued through the ROSES
2014 Astrophysics Data Analysis Program.
The Melbourne group acknowledges support from the Australian Research Council's
Discovery Projects funding scheme (DP150103208).
AvdL is supported by the U.S. Department of Energy under Award Number
DE-SC0018053.
DR is supported by a NASA Postdoctoral Program Senior Fellowship at the NASA
Ames Research Center, administered by the Universities Space Research
Association under contract with NASA.
AS is supported by the ERC-StG `ClustersXCosmo' grant agreement 71676, and by
the FARE-MIUR grant `ClustersXEuclid' R165SBKTMA.
The South Pole Telescope is supported by the National Science Foundation through
grant PLR-1248097. Partial support is also provided by the NSF Physics Frontier
Center grant PHY-1125897 to the Kavli Institute of Cosmological Physics at the
University of Chicago, the Kavli Foundation and the Gordon and Betty Moore
Foundation grant GBMF 947.

\software{\textsc{Astropy} \citep{astropy18}, \textsc{numpy} \citep{numpy},
\textsc{scipy} \citep{scipy}, \textsc{CosmoSIS} \citep{zuntz15},
\textsc{GetDist}\footnote{\url{https://github.com/cmbant/getdist}},
\textsc{matplotlib} \citep{hunter07},
\textsc{pyGTC}\footnote{\url{https://github.com/SebastianBocquet/pygtc}}
\citep{bocquet16b}}

\facilities{South Pole Telescope, Magellan: Clay (Megacam), \hst, \chandra,
Gemini-South (GMOS), PISCO}

\bibliography{spt}

\appendix

\section{Impact of marginalization over the sum of neutrino masses}
\label{sec:appendixneutrino}

\begin{figure}
\epsscale{1.15}
\plottwo{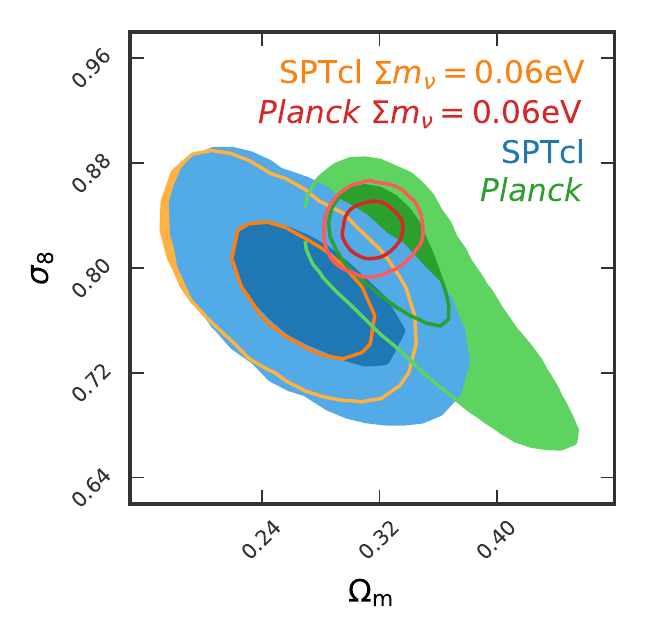}{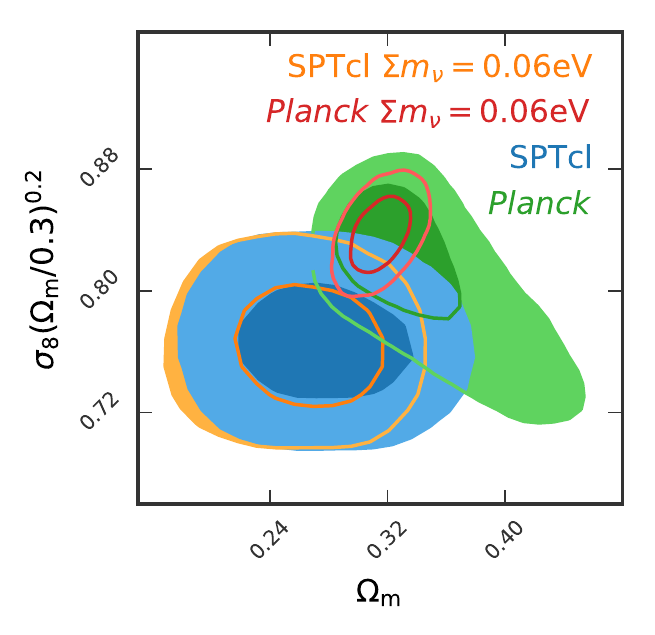}
\caption{Impact of marginalization over \Omnuhh: The cluster constraints are
weakly affected, whereas the \planck\ constraints significantly tighten when
\Omnuhh\ is fixed. Nevertheless, the level of (dis)agreement between these data
sets is not substantially changed by marginalizing over \Omnuhh.}
  \label{fig:LCDM_fix_nu}
\end{figure}

Our baseline analysis is carried out marginalizing over the sum of neutrino
masses (by allowing \Omnuhh\ to vary in the range $0\dots0.01$). In
Fig.~\ref{fig:LCDM_fix_nu}, we show that instead fixing the sum of neutrino
masses to the minimum allowed value from oscillation experiments ($0.06$~eV,
corresponding to $\Omnuhh=6.5\times10^{-4}$) does not qualitatively change the
constraints in the $\Om-\sig$ space from our SPTcl data set. However, as is well
known, the constraints from \planck\ tighten significantly when \Omnuhh\ is
fixed. We note that this tightening does not significantly affect the agreement
between the two probes.

\section{\LCDM\ results: robustness to splits in redshift and impact of X-ray data}
\label{sec:appendixLCDM}

\begin{figure*}
\includegraphics[width=\textwidth]{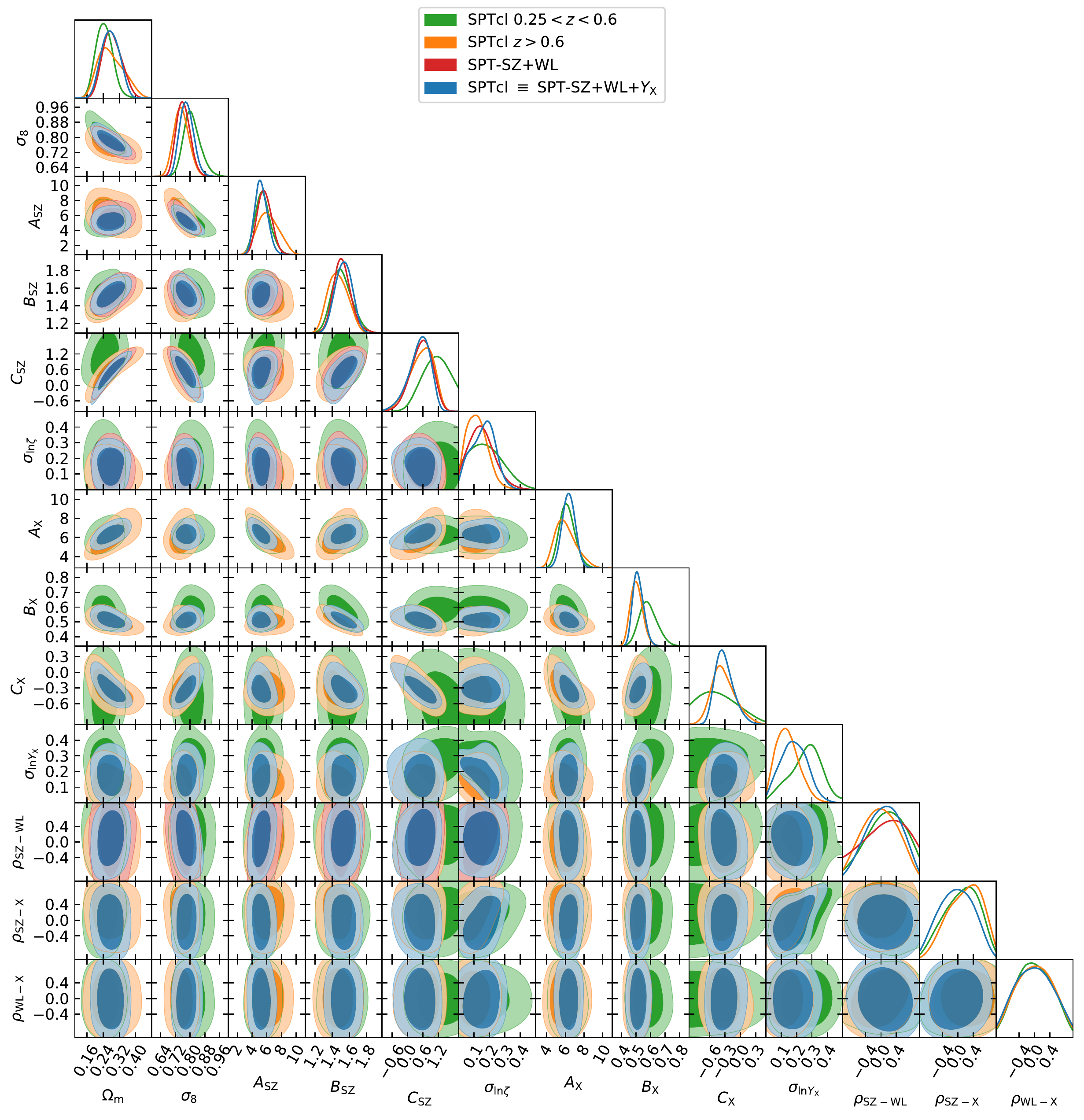}
\caption{\nuLCDM\ constraints on a subset of cosmology and scaling relation
parameters. The full set of fit parameters is listed in
Table~\ref{tab:parameters}. \emph{Blue contours} are obtained from the SPTcl
(SPT-SZ+WL+\yx) dataset, \emph{green contours} are obtained using all clusters
in the redshift range $0.25<z<0.6$, \emph{orange contours} are obtained from the
high-redshift counterpart $z>0.6$, and \emph{red contours} are obtained using
SPT-SZ+WL, without any X-ray data, but with a Gaussian prior applied on
\sigmalnzeta. The inclusion of X-ray does not lead to improved cosmological
constraints, but allows us to drop the prior on scatter \sigmalnzeta\ and to
constrain the X-ray scaling relation. Our current dataset is not able to
constrain any of the correlated scatter coefficients $\rho$. The visual
impression that the $\rho$ parameters are constrained is mostly due to the prior
that the covariance matrix must be positive definite.}
\label{fig:LCDM_WL_Yx_full}
\end{figure*}

Our baseline results are obtained from the SPTcl (SPT-SZ+WL+\yx) data
combination. Here, we show the impact on scaling relation parameters and
cosmology from different cuts. Fig.~\ref{fig:LCDM_WL_Yx_full} shows the most
relevant subset of scaling relation and cosmological parameters for i) the
baseline analysis, ii) an analysis of the low-redshift half of the cluster data
($0.25<z<0.6$), iii) the high-redshift half of the sample ($z>0.6$), iv) the
SPT-SZ+WL data combination, without any X-ray data, but where an informative
Gaussian prior is applied on the SZ scatter ($\sigmalnzeta=\mathcal N(0.13,
0.13^2)$).

Importantly, the cosmological constraints on $\Om-\sig$ are not much affected by
the choice of subsample, and they only vary mildly along the degeneracy axis.

The low-redshift half of the sample only provides weak constraints on the
redshift-evolution of the X-ray scaling relation \cyx. We discussed the
constraints on the X-ray mass-slope \byx\ in Section~\ref{sec:LCDMXray}.

\section{$w$CDM results: robustness to data cuts and impact of X-ray data}
\label{sec:appendixwcdm}

\begin{figure*}
\includegraphics[width=\textwidth]{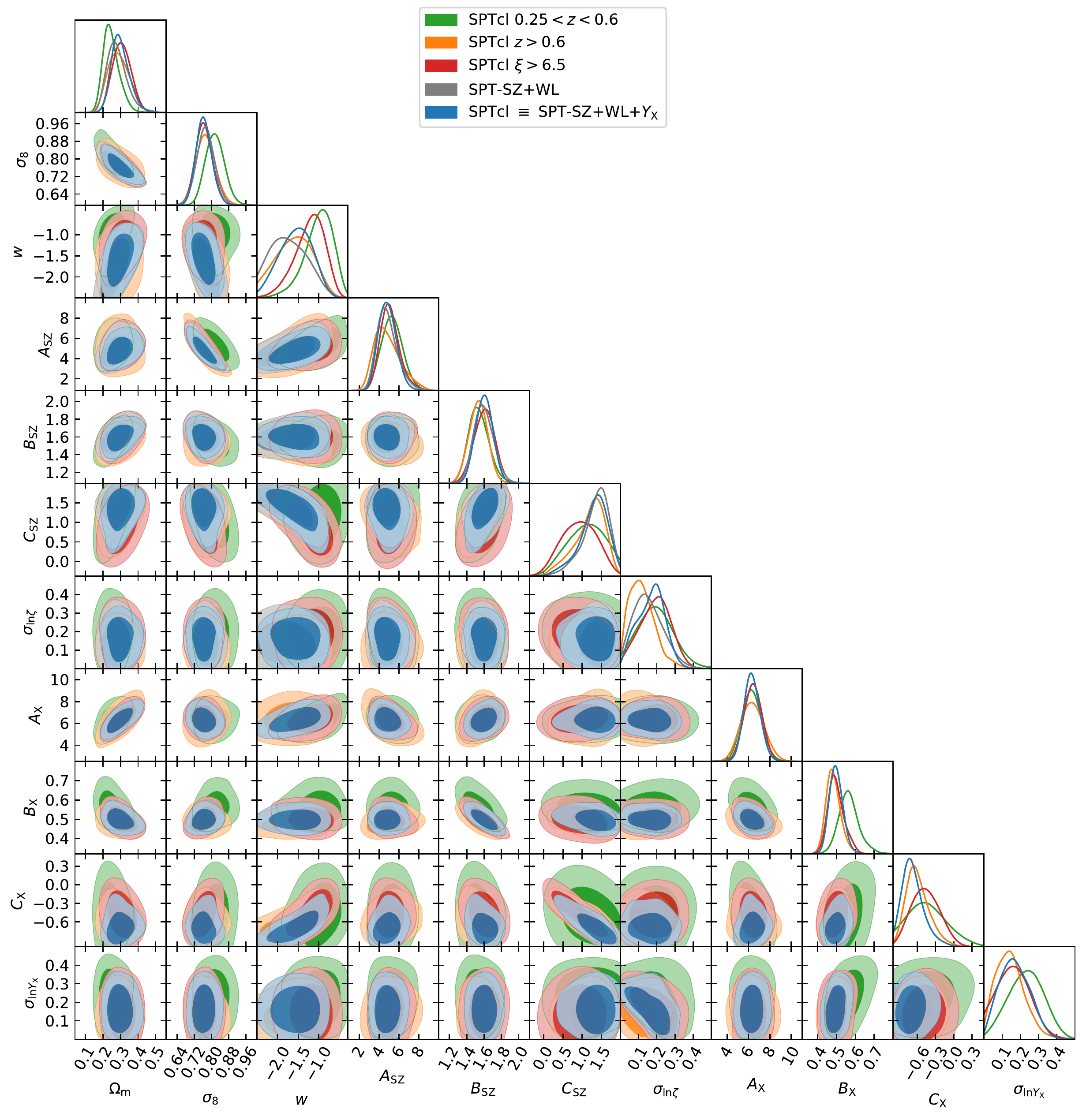}
\caption{\nuwCDM\ constraints on a subset of cosmology and scaling relation
parameters. The full set of fit parameters is listed in
Table~\ref{tab:parameters}. \emph{Blue contours} are obtained from the SPTcl
(SPT-SZ+WL+\yx) dataset, \emph{green contours} are obtained using all clusters
in the redshift range $0.25<z<0.6$, \emph{orange contours} are obtained from the
high-redshift counterpart $z>0.6$, \emph{red contours} are obtained from the
subsample of clusters above SPT detection significance $\xi>6.5$, and \emph{gray
contours} are obtained using SPT-SZ+WL without any X-ray data, but with a
Gaussian prior applied on \sigmalnzeta. Both the low-redshift half of the sample
and the $\xi>6.5$ subsample favor a less negative $w$ in better agreement with a
cosmological constant.}
\label{fig:wCDM_full}
\end{figure*}

As in the previous section, here we discuss the impact of various data cuts on
cosmological constraints, but this time in the context of the \nuwCDM\ model.
Fig.~\ref{fig:wCDM_full} shows the most relevant subset of scaling relation and
cosmological parameters for analyses of i) the baseline cluster sample, ii) the
low-redshift half of the cluster data ($0.25<z<0.6$), iii) the high-redshift
half of the sample ($z>0.6$), iv) a subsample selected above SPT detection
significance $\xi>6.5$, and v) the SPT-SZ+WL data combination, without any X-ray
data, but where an informative Gaussian prior is applied on the SZ scatter
($\sigmalnzeta=\mathcal N(0.13, 0.13^2)$).

As discussed above for the \nuLCDM\ analysis, we see some shifts in the X-ray
slope \byx\ and redshift-evolution \cyx. The constraints on \Om\ and \sig\ are
again not much affected by the choice of subsample. However, while not
statistically significant, there are differences in the recovered values for
$w$. Using the SPT-SZ+WL data combination provides the weakest constraint on
$w$, and its posterior distribution is shifted toward more negative values,
running against the hard prior at $w=-2.5$. Then, as already discussed in
Section~\ref{sec:wCDM}, both the low-redshift half of the sample and the
higher-mass $\xi>6.5$ subsamples prefer slightly higher $w$, with
$w(z<0.6)=\wCDMwloz$ and $w(\xi>6.5)=\wCDMwxigtr$. The high-redshift half of the
sample provides constraints $w(z>0.6)=\wCDMwhiz$ that are very similar to those
from the full sample $w=\wCDMclw$.

\section{The choice of priors: sampling from $A_s$ vs. sampling from $\ln A_s$}

\begin{figure*}
\includegraphics[width=\textwidth]{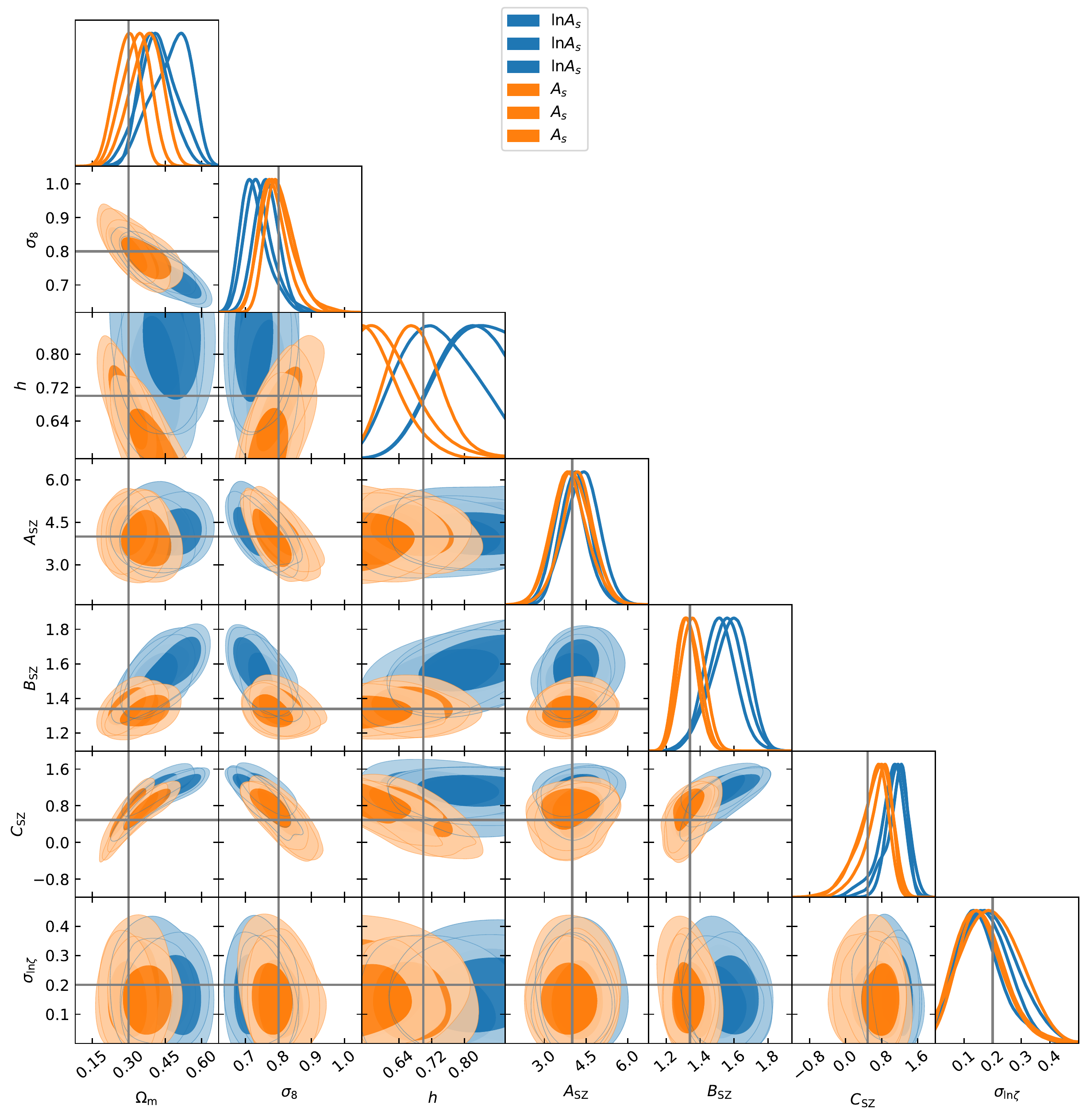}
\caption{Constraints obtained running our analysis pipeline on three
realizations of mock data. For simplicity, the mass calibration is replaced by
Gaussian priors on $\asz=\mathcal N(4, 0.8^2)$ and $\sigmalnzeta=\mathcal
N(0.13, 0.13^2)$. \emph{Blue contours} are obtained sampling from a flat prior
on $\ln(10^{10}A_s)=1\dots4$, \emph{orange contours} are obtained sampling from
a flat prior $A_s=10^{-10}\dots10^{-8}$. The other parameters are sampled from
flat priors. Solid lines show the mock catalog input parameters. Sampling from
$A_s$ performs better in terms of recovering the input values and we choose this
prior throughout this analysis.}
\label{fig:As_lnAs_prior}
\end{figure*}

In this work, we sample from a flat prior on $A_s$, following previous SPT
analyses and e.g., the DES Y1 analysis \citep{des18-main}. In primary CMB
studies however, it is common practice to sample from $\ln 10^{10}A_s$. We test
the impact of this choice of priors by analyzing mock catalogs sampling from a
flat prior on either $A_s$ or $\ln 10^{10}A_s$. Fig.~\ref{fig:As_lnAs_prior}
shows that a flat prior on $A_s$ performs better in terms of recovering the mock
input parameters. This choice of prior does not matter in the limit where $A_s$
(and/or $\ln 10^{10}A_s$) are tightly constrained and we thus expect the impact
of this prior choice to become less important as the constraining power of our
datasets increases. Sampling from a flat prior on \sig\ instead produces results
that are essentially identical to those obtained when sampling from
$\ln10^{10}A_s$.

\newpage
\section{The Cluster Catalog}

\startlongtable
\begin{deluxetable*}{lCC|cc|C|C|CC|C}
\tablewidth{\textwidth}
\tabletypesize{\scriptsize}
\tablecaption{
\label{tab:catalog}
Galaxy cluster candidates with $\xi>4.5$ in the SPT-SZ survey. The subsample
with $\xi\geq5$ and $z\geq0.25$ is used in the cosmological analysis. Clusters
with follow-up WL and/or X-ray data that we use in this work are marked with
``WL'' and/or ``X''. The positions, $\xi$, core radii $\theta_\mathrm{c}$, and
$Y_\mathrm{SZ}$ are the same as presented in \citet{bleem15b}, while the
redshifts marked with * have been updated. Spectroscopic redshifts are quoted
without uncertainties. The mean mass estimates and mass uncertainties take the
intrinsic and measurement scatter into account. We quote redshift lower limits
for unconfirmed SZ detections. The mass estimates $M_\mathrm{500c}$ and
$M_\mathrm{200c}$ are derived from the SPTcl dataset in the \nuLCDM\ model
(Table~\ref{tab:constraints} column 3) and are fully marginalized over cosmology
and scaling relation parameter uncertainties. The estimates
$M_\mathrm{500c}^\mathrm{no~syst.}$ are computed assuming a fixed cosmology and
using the best-fit scaling relation parameters obtained from fitting the SPT-SZ
number counts against that fixed cosmology \citep[this approach was also adopted
in][] {bleem15b}. The catalog can also be found at
\url{https://pole.uchicago.edu/public/data/sptsz-clusters}.}
\tablehead{
  \colhead{SPT~ID} & \colhead{R.A.} & \colhead{DEC.} &
  \dcolhead{\xi} &  \dcolhead{\theta_\mathrm{c}} & \dcolhead{Y_\mathrm{SZ}} &
  \colhead{Redshift} &
  \dcolhead{M_\mathrm{500c}} &
  \dcolhead{M_\mathrm{200c}} &
  \dcolhead{M_\mathrm{500c}^\mathrm{no~ syst.}} \\
  \colhead{} & \colhead{(J2000)} & \colhead{(J2000)} & \colhead{} & \colhead{arcmin} & \dcolhead{10^{-6}\mathrm{arcmin}^2} & \colhead{} & \dcolhead{[10^{14}M_\odot/h_{70}]} & \dcolhead{[10^{14}M_\odot/h_{70}]}
  }
\startdata
\input{catalog}
\enddata
\end{deluxetable*}

\end{document}

%% file: results.tex
\newcommand{\LCDMCe}{\ensuremath{439.8}}
\newcommand{\LCDMCv}{\ensuremath{26.8}}
\newcommand{\LCDMCd}{\ensuremath{449.3}}

\newcommand{\wCDMwYx}{\ensuremath{-1.53^{+0.36}_{-0.25}}}

\newcommand{\coupledrho}{\ensuremath{0.55-0.60}}
\newcommand{\decoupledrho}{\ensuremath{0.06-0.12}}

\newcommand{\LCDMclOm}{\ensuremath{0.276\pm0.047}}
\newcommand{\LCDMclsig}{\ensuremath{0.781\pm0.037}}
\newcommand{\LCDMclsigOmtwo}{\ensuremath{0.766\pm0.025}}

\newcommand{\LCDMclAyx}{\ensuremath{6.35\pm0.69}}
\newcommand{\LCDMclByx}{\ensuremath{0.514\pm0.037}}
\newcommand{\LCDMclCyx}{\ensuremath{-0.31^{+0.14}_{-0.21}}}
\newcommand{\LCDMclsigmalnyx}{\ensuremath{0.18\pm0.09}}
\newcommand{\LCDMByxloz}{\ensuremath{0.583^{+0.054}_{-0.069}}}
\newcommand{\LCDMByxhiz}{\ensuremath{0.503^{+0.037}_{-0.047}}}
\newcommand{\LCDMAmg}{\ensuremath{0.116\pm0.011}}
\newcommand{\LCDMBmg}{\ensuremath{1.22\pm0.07}}
\newcommand{\LCDMBmgloz}{\ensuremath{1.12\pm0.09}}
\newcommand{\LCDMBmghiz}{\ensuremath{1.36\pm0.11}}
\newcommand{\LCDMCmg}{\ensuremath{-0.05\pm0.17}}
\newcommand{\LCDMDmg}{\ensuremath{0.11\pm0.04}}

\newcommand{\LCDMPlancksigOmtwo}{\ensuremath{0.814^{+0.041}_{-0.020}}}

\newcommand{\LCDMclPlOm}{\ensuremath{0.353\pm0.027}}
\newcommand{\LCDMclPlsig}{\ensuremath{0.761\pm0.033}}
\newcommand{\LCDMclPlsigOmtwo}{\ensuremath{0.786\pm0.025}}

\newcommand{\LCDMclPlmnu}{\ensuremath{0.39\pm0.19}}
\newcommand{\LCDMclPlmnuNinetyfive}{\ensuremath{0.74}}

\newcommand{\LCDMclPlTTtaumnu}{\ensuremath{0.35\pm0.21}}
\newcommand{\LCDMclzsmpsixPlmnu}{\ensuremath{0.29^{+0.09}_{-0.29}}}

\newcommand{\wCDMclw}{\ensuremath{-1.55\pm0.41}}
\newcommand{\wCDMclOm}{\ensuremath{0.299\pm0.049}}
\newcommand{\wCDMclsig}{\ensuremath{0.766\pm0.036}}

\newcommand{\wCDMwloz}{\ensuremath{-1.01^{+0.41}_{-0.25}}}
\newcommand{\wCDMwhiz}{\ensuremath{-1.58\pm0.46}}
\newcommand{\wCDMwxigtr}{\ensuremath{-1.21^{+0.42}_{-0.29}}}

\newcommand{\wCDMclPlOm}{\ensuremath{0.347\pm0.039}}
\newcommand{\wCDMclPlsig}{\ensuremath{0.761\pm0.027}}
\newcommand{\wCDMclPlw}{\ensuremath{-1.12\pm0.21}}
\newcommand{\wCDMclPlmnu}{\ensuremath{0.50\pm0.24}}

\newcommand{\Byxpvalue}{\ensuremath{0.021}}
\newcommand{\Byxsigma}{\ensuremath{2.3}}
\newcommand{\LCDMBmgsigma}{\ensuremath{2.5\sigma}}

\newcommand{\wCDMpvalue}{\ensuremath{0.076}}
\newcommand{\wCDMsigma}{\ensuremath{1.8}}
\newcommand{\LCDMcorrAszMnu}{\ensuremath{0.83}}
\newcommand{\nuLCDMsummnusigma}{\ensuremath{2.0}}
\newcommand{\nuLCDMtausummnusigma}{\ensuremath{1.7}}

\newcommand{\LCDMclPlimproveOm}{\ensuremath{3}}
\newcommand{\LCDMclPlimprovesig}{\ensuremath{12}}
\newcommand{\LCDMclPlimprovesigOmtwo}{\ensuremath{20}}

\newcommand{\mnuupperlimitshift}{\ensuremath{63}}
\newcommand{\nuwCDMshiftsig}{\ensuremath{-0.031}}

\newcommand{\massshiftL}{\ensuremath{8}}
\newcommand{\massshiftw}{\ensuremath{4}}

\newcommand{\LCDMByxlohipvalue}{\ensuremath{0.44}}
\newcommand{\LCDMByxlohisigma}{\ensuremath{0.8}}
\newcommand{\nuLCDMclPlpvalue}{\ensuremath{0.13}}
\newcommand{\nuLCDMclPlsigma}{\ensuremath{1.5}}
\newcommand{\nuLCDMclPlsigOmtwopvalue}{\ensuremath{0.28}}
\newcommand{\nuLCDMclPlsigOmtwosigma}{\ensuremath{1.1}}
\newcommand{\dHpvalue}{\ensuremath{0.86}}
\newcommand{\dHsigma}{\ensuremath{0.2}}

%% file: authors.tex
\author{S.~Bocquet}
\affiliation{\Munich}
\affiliation{\Argonne}
\affiliation{\KICPChicago}

\author{J.~P.~Dietrich}
\affiliation{\Munich}
\affiliation{\ExcellenceCluster}

\author{T.~Schrabback}
\affiliation{\AIfA}

\author{L.~E.~Bleem}
\affiliation{\Argonne}
\affiliation{\KICPChicago}

\author{M.~Klein}
\affiliation{\Munich}
\affiliation{\MPE}

\author{S.~W.~Allen}
\affiliation{\StanfordKPAC}
\affiliation{\StanfordPhys}
\affiliation{\SLAC}

\author{D.~E.~Applegate}
\affiliation{\KICPChicago}

\author{M.~L.~N.~Ashby}
\affiliation{\CfA}

\author{M.~Bautz}
\affiliation{\MIT}

\author{M.~Bayliss}
\affiliation{\MIT}

\author{B.~A.~Benson}
\affiliation{\AAUChicago}
\affiliation{\KICPChicago}
\affiliation{\FNAL}

\author{M.~Brodwin}
\affiliation{\Miss}

\author{E.~Bulbul}
\affiliation{\CfA}

\author{R.~E.~A.~Canning}
\affiliation{\StanfordKPAC}
\affiliation{\StanfordPhys}

\author{R.~Capasso}
\affiliation{\Munich}
\affiliation{\ExcellenceCluster}

\author{J.~E.~Carlstrom}
\affiliation{\AAUChicago}
\affiliation{\KICPChicago}
\affiliation{\PhysicsUChicago}
\affiliation{\Argonne}
\affiliation{\EFIChicago}

\author{C.~L.~Chang}
\affiliation{\AAUChicago}
\affiliation{\KICPChicago}
\affiliation{\Argonne}

\author{I.~Chiu}
\affiliation{\ASIAA}

\author{H-M.~Cho}
\affiliation{\NIST}

\author{A.~Clocchiatti}
\affiliation{\PUC}

\author{T.~M.~Crawford}
\affiliation{\AAUChicago}
\affiliation{\KICPChicago}

\author{A.~T.~Crites}
\affiliation{\AAUChicago}
\affiliation{\KICPChicago}
\affiliation{\Caltech}

\author{T.~de~Haan}
\affiliation{\Berkeley}

\author{S.~Desai}
\affiliation{\Hyderabad}

\author{M.~A.~Dobbs}
\affiliation{\McGill}
\affiliation{\CIFAR}

\author{R.~J.~Foley}
\affiliation{\illast}
\affiliation{\illphy}

\author{W.~R.~Forman}
\affiliation{\CfA}

\author{G.~P.~Garmire}
\affiliation{\Huntingdon}

\author{E.~M.~George}
\affiliation{\Berkeley}
\affiliation{\MPE}

\author{M.~D.~Gladders}
\affiliation{\AAUChicago}
\affiliation{\KICPChicago}

\author{A.~H.~Gonzalez}
\affiliation{\UFlorida}

\author{S.~Grandis}
\affiliation{\Munich}
\affiliation{\ExcellenceCluster}

\author{N.~Gupta}
\affiliation{\Melbourne}

\author{N.~W.~Halverson}
\affiliation{\Colorado}

\author{J.~Hlavacek-Larrondo}
\affiliation{\UMon}
\affiliation{\KIPAC}
\affiliation{\StanfordPhys}

\author{H.~Hoekstra}
\affiliation{\LeidenObservatory}

\author{G.~P.~Holder}
\affiliation{\McGill}

\author{W.~L.~Holzapfel}
\affiliation{\Berkeley}

\author{Z.~Hou}
\affiliation{\KICPChicago}
\affiliation{\PhysicsUChicago}

\author{J.~D.~Hrubes}
\affiliation{\UChicago}

\author{N.~Huang}
\affiliation{\Berkeley}

\author{C.~Jones}
\affiliation{\CfA}

\author{G.~Khullar}
\affiliation{\KICPChicago}
\affiliation{\AAUChicago}

\author{L.~Knox}
\affiliation{\Davis}

\author{R.~Kraft}
\affiliation{\CfA}

\author{A.~T.~Lee}
\affiliation{\Berkeley}
\affiliation{\LBNL}

\author{A.~von~der~Linden}
\affiliation{\StonyBrook}

\author{D.~Luong-Van}
\affiliation{\UChicago}

\author{A.~Mantz}
\affiliation{\StanfordKPAC}
\affiliation{\StanfordPhys}

\author{D.~P.~Marrone}
\affiliation{\Arizona}

\author{M.~McDonald}
\affiliation{\MIT}

\author{J.~J.~McMahon}
\affiliation{\Michigan}

\author{S.~S.~Meyer}
\affiliation{\AAUChicago}
\affiliation{\KICPChicago}
\affiliation{\PhysicsUChicago}
\affiliation{\EFIChicago}

\author{L.~M.~Mocanu}
\affiliation{\AAUChicago}
\affiliation{\KICPChicago}

\author{J.~J.~Mohr}
\affiliation{\Munich}
\affiliation{\MPE}
\affiliation{\ExcellenceCluster}

\author{R.~G.~Morris}
\affiliation{\SLAC}
\affiliation{\StanfordKPAC}

\author{S.~Padin}
\affiliation{\AAUChicago}
\affiliation{\KICPChicago}
\affiliation{\Caltech}

\author{S.~Patil}
\affiliation{\Melbourne}

\author{C.~Pryke}
\affiliation{\Minnesota}

\author{D.~Rapetti}
\affiliation{\Munich}
\affiliation{\ExcellenceCluster}
\affiliation{\Colorado}
\affiliation{\NASAAMes}

\author{C.~L.~Reichardt}
\affiliation{\Melbourne}

\author{A.~Rest}
\affiliation{\STScI}

\author{J.~E.~Ruhl}
\affiliation{\CaseWestern}

\author{B.~R.~Saliwanchik}
\affiliation{\Yale}

\author{A.~Saro}
\affiliation{\Trieste}
\affiliation{\INAF}
\affiliation{\IFPU}
\affiliation{\Munich}
\affiliation{\ExcellenceCluster}

\author{J.~T.~Sayre}
\affiliation{\Colorado}

\author{K.~K.~Schaffer}
\affiliation{\KICPChicago}
\affiliation{\EFIChicago}
\affiliation{\ArtInstChicago}

\author{E.~Shirokoff}
\affiliation{\AAUChicago}
\affiliation{\KICPChicago}

\author{B.~Stalder}
\affiliation{\CfA}

\author{S.~A.~Stanford}
\affiliation{\Davis}
\affiliation{\LLNL}

\author{Z.~Staniszewski}
\affiliation{\CaseWestern}

\author{A.~A.~Stark}
\affiliation{\CfA}

\author{K.~T.~Story}
\affiliation{\StanfordKPAC}
\affiliation{\StanfordPhys}

\author{V.~Strazzullo}
\affiliation{\Munich}

\author{C.~W.~Stubbs}
\affiliation{\CfA}
\affiliation{\Harvard}

\author{K.~Vanderlinde}
\affiliation{\Dunlap}
\affiliation{\Toronto}

\author{J.~D.~Vieira}
\affiliation{\illast}
\affiliation{\illphy}

\author{A. Vikhlinin}
\affiliation{\CfA}

\author{R.~Williamson}
\affiliation{\AAUChicago}
\affiliation{\KICPChicago}
\affiliation{\Caltech}

\author{A.~Zenteno}
\affiliation{\CTIO}


\def\AAUChicago{Department of Astronomy and Astrophysics, University of Chicago, Chicago, IL 60637, USA}
\def\AIfA{Argelander-Institut f\"ur Astronomie, Auf dem H\"ugel 71, 53121 Bonn, Germany}
\def\Argonne{HEP Division, Argonne National Laboratory, Argonne, IL 60439, USA}
\def\Arizona{Steward Observatory, University of Arizona, 933 North Cherry Avenue, Tucson, AZ 85721, USA}
\def\ArtInstChicago{Liberal Arts Department, School of the Art Institute of Chicago, Chicago, IL 60603, USA}
\def\ASIAA{Institute of Astronomy and Astrophysics, Academia Sinica, Taipei 10617, Taiwan}
\def\Berkeley{Department of Physics, University of California, Berkeley, CA 94720, USA}
\def\Caltech{California Institute of Technology, Pasadena, CA 91125, USA}
\def\CaseWestern{Physics Department, Center for Education and Research in Cosmology and Astrophysics, Case Western Reserve University, Cleveland, OH 44106, USA}
\def\CfA{Center for Astrophysics \textbar\ Harvard \& Smithsonian, Cambridge MA 02138, USA}
\def\CIFAR{Canadian Institute for Advanced Research, CIFAR Program in Cosmology and Gravity, Toronto, ON, M5G 1Z8, Canada}
\def\Colorado{Department of Astrophysical and Planetary Sciences and Department of Physics, University of Colorado, Boulder, CO 80309, USA}
\def\CTIO{Cerro Tololo Inter-American Observatory, Casilla 603, La Serena, Chile}
\def\DARK{Dark Cosmology Centre, Niels Bohr Institute, University of Copenhagen Juliane Maries Vej 30, 2100 Copenhagen, Denmark}
\def\Davis{Department of Physics, University of California, Davis, CA 95616, USA}
\def\Dunlap{Dunlap Institute for Astronomy \& Astrophysics, University of Toronto, 50 St George St, Toronto, ON, M5S 3H4, Canada}
\def\EFIChicago{Enrico Fermi Institute, University of Chicago, Chicago, IL 60637, USA}
\def\ExcellenceCluster{Excellence Cluster Universe, Boltzmannstr.\ 2, 85748 Garching, Germany}
\def\FNAL{Fermi National Accelerator Laboratory, Batavia, IL 60510-0500, USA}
\def\Harvard{Department of Physics, Harvard University, 17 Oxford Street, Cambridge, MA 02138, USA}
\def\Huntingdon{Huntingdon Institute for X-ray Astronomy, LLC, USA}
\def\Hyderabad{Department of Physics, IIT Hyderabad, Kandi, Telangana 502285, India}
\def\IFPU{Institute for Fundamental Physics of the Universe, Via Beirut 2, 34014 Trieste, Italy}
\def\illast{Astronomy Department, University of Illinois at Urbana-Champaign, 1002 W.\ Green Street, Urbana, IL 61801, USA}
\def\illphy{Department of Physics, University of Illinois Urbana-Champaign, 1110 W.\ Green Street, Urbana, IL 61801, USA}
\def\INAF{INAF-Osservatorio Astronomico di Trieste, via G. B. Tiepolo 11, 34143 Trieste, Italy}
\def\KICPChicago{Kavli Institute for Cosmological Physics, University of Chicago, Chicago, IL 60637, USA}
\def\KIPAC{Kavli Institute for Particle Astrophysics and Cosmology, Stanford University, 452 Lomita Mall, Stanford, CA 94305-4085, USA}
\def\LBNL{Physics Division, Lawrence Berkeley National Laboratory, Berkeley, CA 94720, USA}
\def\LeidenObservatory{Leiden Observatory, Leiden University, Niels Bohrweg 2, 2333 CA, Leiden, the Netherlands}
\def\LLNL{Institute of Geophysics and Planetary Physics, Lawrence Livermore National Laboratory, Livermore, CA 94551, USA}
\def\McGill{Department of Physics, McGill University, Montreal, Quebec H3A 2T8, Canada}
\def\Melbourne{School of Physics, University of Melbourne, Parkville, VIC 3010, Australia}
\def\Michigan{Department of Physics, University of Michigan, Ann Arbor, MI 48109, USA}
\def\Minnesota{Department of Physics, University of Minnesota, Minneapolis, MN 55455, USA}
\def\Miss{Department of Physics and Astronomy, University of Missouri, 5110 Rockhill Road, Kansas City, MO 64110, USA}
\def\MIT{Kavli Institute for Astrophysics and Space Research, Massachusetts Institute of Technology, 77 Massachusetts Avenue, Cambridge, MA~02139, USA}
\def\MPE{Max Planck Institute for Extraterrestrial Physics, Giessenbachstr.\ 1, 85748 Garching, Germany}
\def\Munich{Faculty of Physics, Ludwig-Maximilians-Universit\"{a}t, Scheinerstr.\ 1, 81679 Munich, Germany}
\def\NASAAMes{NASA Ames Research Center, Moffett Field, CA 94035, USA}
\def\NIST{NIST Quantum Devices Group, Boulder, CO 80305, USA}
\def\PhysicsUChicago{Department of Physics, University of Chicago, Chicago, IL 60637, USA}
\def\PUC{Departamento de Astronomia y Astrosifica, Pontificia Universidad Catolica, Chile}
\def\SLAC{SLAC National Accelerator Laboratory, 2575 Sand Hill Road, Menlo Park, CA 94025, USA}
\def\StanfordKPAC{Kavli Institute for Particle Astrophysics and Cosmology, Stanford University, 452 Lomita Mall, Stanford, CA 94305, USA}
\def\StanfordPhys{Department of Physics, Stanford University, 382 Via Pueblo Mall, Stanford, CA 94305, USA}
\def\StonyBrook{Department of Physics and Astronomy, Stony Brook University, Stony Brook, NY 11794, USA}
\def\STScI{Space Telescope Science Institute, 3700 San Martin Dr., Baltimore, MD 21218, USA}
\def\Toronto{Department of Astronomy \& Astrophysics, University of Toronto, 50 St George St, Toronto, ON, M5S 3H4, Canada}
\def\Trieste{Astronomy Unit, Department of Physics, University of Trieste, via Tiepolo 11, 34131 Trieste, Italy}
\def\UChicago{University of Chicago, Chicago, IL 60637, USA}
\def\UFlorida{Department of Astronomy, University of Florida, Gainesville, FL 32611, USA}
\def\UMon{Department of Physics, Universit\'e de Montr\'eal, Montreal, Quebec H3T 1J4, Canada}
\def\Yale{Department of Physics, Yale University, New Haven, CT 06511, USA}

%% file: abstract.tex
We derive cosmological constraints using a galaxy cluster sample selected from
the 2500~deg$^2$ SPT-SZ survey. The sample spans the redshift range $0.25<
z<1.75$ and contains 343 clusters with SZ detection significance $\xi>5$. The
sample is supplemented with optical weak gravitational lensing measurements of
32 clusters with $0.29<z<1.13$ (from Magellan and HST) and X-ray measurements of
89 clusters with $0.25<z<1.75$ (from \textit{Chandra}). We rely on minimal
modeling assumptions: i) weak lensing provides an accurate means of measuring
halo masses, ii) the mean SZ and X-ray observables are related to the true halo
mass through power-law relations in mass and dimensionless Hubble parameter
$E(z)$ with a-priori unknown parameters, iii) there is (correlated, lognormal)
intrinsic scatter and measurement noise relating these observables to their mean
relations. We simultaneously fit for these astrophysical modeling parameters and
for cosmology. Assuming a flat $\nu\Lambda$CDM model, in which the sum of
neutrino masses is a free parameter, we measure
$\Omega_\mathrm{m}=0.276\pm0.047$, $\sigma_8=0.781\pm0.037$, and
$\sigma_8(\Omega_\mathrm{m}/0.3)^{0.2}=0.766\pm0.025$. The redshift evolution of
the X-ray $Y_\mathrm{X}$--mass and $M_\mathrm{gas}$--mass relations are both
consistent with self-similar evolution to within $1\sigma$. The mass-slope of
the $Y_\mathrm{X}$--mass relation shows a $2.3\sigma$ deviation from
self-similarity. Similarly, the mass-slope of the $M_\mathrm{gas}$--mass
relation is steeper than self-similarity at the $2.5\sigma$ level. In a $\nu
w$CDM cosmology, we measure the dark energy equation of state parameter
$w=-1.55\pm0.41$ from the cluster data. We perform a measurement of the growth
of structure since redshift $z\sim1.7$ and find no evidence for tension with the
prediction from General Relativity. This is the first analysis of the SPT
cluster sample that uses direct weak-lensing mass calibration, and is a step
toward using the much larger weak-lensing dataset from DES. We provide updated
redshift and mass estimates for the SPT sample.

%% file: catalog.tex
SPT-CL~J0000-4356 & 0.0663 & -43.9494  & 5.92 & 0.25 & 77\pm21 & 0.73\pm0.05^* & 3.55^{+0.65}_{-0.83} & 5.56^{+1.04}_{-1.34} & 4.01^{+0.67}_{-0.74} \\
SPT-CL~J0000-5748$^\text{WL, X}$ & 0.2499 & -57.8064  & 8.49 & 0.50 & 83\pm13 & 0.702 & 4.33^{+0.65}_{-0.86} & 6.79^{+1.04}_{-1.37} & 4.88^{+0.59}_{-0.71} \\
SPT-CL~J0001-4024 & 0.3610 & -40.4108  & 5.42 & 0.75 & 60\pm13 & 0.83\pm0.03^* & 3.19^{+0.65}_{-0.83} & 5.03^{+1.04}_{-1.33} & 3.62^{+0.64}_{-0.73} \\
SPT-CL~J0001-4842 & 0.2768 & -48.7132  & 5.69 & 1.25 & 78\pm13 & 0.33\pm0.02^* & 3.89^{+0.75}_{-0.90} & 5.92^{+1.16}_{-1.40} & 4.48^{+0.75}_{-0.83} \\
SPT-CL~J0001-5440 & 0.4059 & -54.6697  & 5.69 & 1.00 & 60\pm13 & 0.82\pm0.08^* & 3.47^{+0.64}_{-0.88} & 5.46^{+1.02}_{-1.42} & 3.88^{+0.67}_{-0.75} \\
SPT-CL~J0001-6258 & 0.4029 & -62.9808  & 4.69 & 1.50 & 52\pm16 & 0.21\pm0.02 & 3.32^{+0.66}_{-0.88} & 4.97^{+1.00}_{-1.34} & 3.89^{+0.65}_{-0.87} \\
SPT-CL~J0002-5224 & 0.6433 & -52.4092  & 4.67 & 1.00 & 48\pm13 & >0.71 & \nodata & \nodata & \nodata \\
SPT-CL~J0002-5557 & 0.5138 & -55.9621  & 5.20 & 0.25 & 55\pm18 & 1.15\pm0.10 & 2.84^{+0.58}_{-0.81} & 4.55^{+0.95}_{-1.32} & 3.20^{+0.56}_{-0.68} \\